\documentclass[%
 preprint,
 amsmath,amssymb,
 aps,
]{revtex4-1}

\usepackage{amsmath}
\usepackage{graphicx,epstopdf}
\usepackage{dcolumn}
\usepackage{bm}
\usepackage{xcolor}
\usepackage{float}
\usepackage{enumerate}
\usepackage{natbib}
\usepackage{multirow}
\usepackage{hyperref}

\begin{document}

\title{Phonon hydrodynamics in crystalline materials}

\author{Kanka Ghosh}
\email{kanka.ghosh@u-bordeaux.fr}
\affiliation{University of Bordeaux, I2M Laboratory, UMR CNRS 5295, 351 Cours de la libération, F-33400 Talence, France}
\author{Andrzej Kusiak}%
\affiliation{University of Bordeaux, I2M Laboratory, UMR CNRS 5295, 351 Cours de la libération, F-33400 Talence, France}
 \author{Jean-Luc Battaglia}%
 \affiliation{University of Bordeaux, I2M Laboratory, UMR CNRS 5295, 351 Cours de la libération, F-33400 Talence, France}


\begin{abstract}

Phonon hydrodynamics is an exotic phonon transport phenomenon that challenges the conventional understanding of diffusive phonon scattering in crystalline solids. It features a peculiar collective motion of phonons with various unconventional properties resembling fluid hydrodynamics, facilitating non Fourier heat transport. Hence, it opens up several new avenues to enrich the knowledge and implementations on phonon physics, phonon engineering, and micro and nanoelectronic device technologies. This review aims at covering a comprehensive development as well as the recent advancements in this field via experiments, analytical methods, and state-of-the-art numerical techniques. The evolution of the topic has been realized using both phenomenological and material science perspectives. Further, the discussions related to the factors that influence such peculiar motion, illustrate the capability of phonon hydrodynamics to be implemented in various applications. A plethora of new ideas can emerge from the topic considering both the physics and the material science axes, navigating towards a promising outlook in the research areas around phonon transport in non-metallic solids.
 
\end{abstract}

\maketitle


\section{Introduction:}

Phonons are quasi-particles, which are primarily hold responsible for the transport of heat in non-metallic solids. The effect of phonons in heat transport, is realized via thermal conductivity that bears significant importance in characterization, engineering and applications of heat transport in solids \cite{ChenReviewphonon2021NatMat}. Solving various problems related to heat transport are inevitable to our daily lives as well as to the future technological advancements. These include thermal management of various devices, thermal characterization of different electronic, photonic and phononic materials, thermal insulation, energy conversion, high temperature applications of devices and what not \cite{ChenReviewphonon2021NatMat}. Manipulation of phononic properties of materials via advanced state-of-the-art experimental and theoretical techniques enables achieving such applications with great flexibility \cite{chen2018manipulation, kimphononreviez2021}. In this context, it is important to point out that both high and low thermal conductivity materials are equally important to solve distinct problems related to the heat transport in solids. While high thermal conductivity materials help discovering applications in the domain of thermal dissipation in microelectronics (e.g. the usage of graphene as an efficient heat spreading material in high power driven systems \cite{wanggraphenesmall2018} due to its enormous thermal conductivity of $\approx$ 3000 - 5000 W/mK at room temperature \cite{Balandingraphene}), low thermal conductivity materials also help developing efficient applications in thermoelectrics (e.g. single crystal PbTe with an extremely small thermal conductivity of around 2.2 W/mK at room temperature, used as one of the best thermoelectric materials \cite{RomeroPbTe2015, JuPbTe2018}).

The process of heat transfer in a solid has been traditionally understood by the celebrated Fourier's law of heat conduction, where phonons are treated to scatter diffusively. Thus the thermal gradient developed across the solid and the resulting heat current density are connected via the thermal transport coefficient named thermal conductivity, describing the phenomena of heat conduction in solid, via 
\begin{equation}
    \textbf{Q} (\textbf{x}, t) = -\kappa \nabla T (\textbf{x}, t)
\end{equation}
where $T (\textbf{x}, t)$ is the local temperature field, $Q (\textbf{x}, t)$ is the local heat current density and $\kappa$ denotes the thermal conductivity. Connecting with energy density ($e$), the Fourier's law leads to the continuity equation 
\begin{equation}
    \nabla \cdot \textbf{Q} + \frac{\partial e}{\partial t} = 0
\end{equation}
\noindent 
However, the Fourier's law breaks down in certain situations that give rise to peculiar, anomalous and exotic phenomena related to heat conduction \cite{GangChenReview2021, changfourierlawbreak2008, dhar2019fP, dhar2008, Physrep2015Guo}. Phonon hydrodynamics \cite{Gurzhi1968, BeckReview1974,  Broido, Ackerman1966solidhelium, Cepellotti2015, Leebookchapter2020, Lindsay2019, YuReview2021} is one such phenomena where phonons flow collectively instead of diffusively. This causes a surge in the thermal conductivity up to infinite unless thermal resistance starts acting against the phonon flow. Several similarities are drawn between this peculiar flow of phonons with that of the fluids. Firstly, in fluid hydrodynamics, collective motion of fluids are important rather than the motion of individual atoms constituting the fluid. Similarly, phonon hydrodynamics addresses the collective, coherent flow of phonons. Both the fluid flow and phonon flow can be well described by the Boltzmann transport equation considering the distribution of particles in fluids and phonons in solids. Temperature gradient serves as a driving force for the phonon flow in phonon hydrodynamics. Likewise in the macroscopic picture of fluid flow, pressure gradient drives the fluid molecules which is described by Euler's equation \cite{Leebookchapter2020} (or in more general picture by the Navier-Stokes equation).

Nevertheless, some microscopic characteristics and manifestations of fluids and phonons are also notably different in terms of hydrodynamics. For example, in fluid flow, total momentum is always conserved upon scattering between the constituting atoms while depending on the scattering mechanism between them, phonons can perform either momentum conserving or momentum destroying events \cite{Chen2018R, Leebookchapter2020}. The momentum conserving and destroying scattering events are called `Normal' (N) and `Umklapp' (U) scattering respectively where N scattering assists the collective motion and U scattering impedes the flow, enabling the thermal resistance to the phonon flow \cite{Callaway1959}. These scattering events, along with grain boundaries and impurities in the crystal, lead to distinct phonon transport regimes in crystalline solids as a function of temperature \cite{kaviany_2014, Cepellotti2015}. At low temperatures, phonon mean free paths are much larger than the characteristic length of the sample which help phonons to propagate ballistically towards the boundaries, indicating a system size dependent thermal transport in the ballistic regime. At the other extreme of high temperature, phonons possess smaller mean free paths compared to the size of the crystal and they scatter diffusively, vanishing completely the size dependency in the thermal transport properties. This diffusive thermal transport regime is dominated by U scattering events where phonons with large wave vectors scatter with each other causing a non conservation of phonon momentum and reversal of the direction of phonon propagation. The thermal transport regime, in between theses two extremes, comprises an intermediate temperature window where small wave vectors excite and phonons dominantly perform momentum conserving N scattering that leads to hydrodynamic phonon flow. By the very nature of this transport regime it emerges as a fragile and hard to achieve in most of the solids because of the simultaneous strict requirements of both less U scattering and high N scattering events \cite{Lindsay2019, Leebookchapter2020, GK2:1966}. Lowering of temperature helps achieving in exciting only small wave vectors to avoid U scattering but also affects the frequent occurrence of many N scattering events. 

The inception of various advanced experimental and first-principle based methods, helped discovering a larger pool of materials having either high Debye temperature (to postpone the U scattering at later temperature) or large anharmonicity (to induce more N scattering events) which is difficult to observe in same material. Further, these advanced methods assist in identifying the controlling parameters and therefore pave the way towards phonon-engineering to uncover new possibilities of applications of these materials. Some of the two dimensional materials (e.g. graphene) had been found \cite{Cepellotti2015, Broido, YuReview2021} to exhibit strong N scattering even at room temperature owing to their out-of-plane flexural acoustic modes. Some isotopically not so pure materials (e.g. SrTiO$_3$) had been experimentally found \cite{MartelliStrontium2018, Koreeda2009SrTio3} to feature strong anharmonicity and therefore strong N scattering due to the presence of soft optical modes. One dimensional single walled carbon nanotubes had also been found \cite{Lee2017} to possess phonon hydrodynamics. Recent experiments on graphite had been shown \cite{Ding2022, Huberman2019graphite} to feature phonon hydrodynamics even above a temperature of 200 K. Apart from envisaging more accurate description of phonon thermal transport as a function of temperature, all these realizations can drive the emergence of a lot of interesting heat transport applications keeping phonon hydrodynamics as a focal point. For example, graphene having a prominent presence of phonon hydrodynamics up to a fairly high temperature (up to 300 K \cite{Cepellotti2015}), can be used in the applications of thermal rectification and thermal signal transmitters \cite{Physrep2015Guo, Broido}.

This review is structured as follows: Section \ref{sec:2} approaches the idea of phonon hydrodynamics from a straightforward scattering rate analysis starting from describing various phonon scattering processes. Section \ref{sec:3} discusses the onset of phonon hydrodynamics from a new `relaxon' perspective in approaching the phonon hydrodynamics. Looking from a phenomenological point of view, several important features of phonon hydrodynamics have been thoroughly explored and explained in section \ref{sec:4}, examining over distinct theoretical, experimental and numerical efforts. Analyzing from a material scientist's viewpoint, section \ref{sec:5} scrutinizes a detailed and up-to-date account of different 3D, 2D and 1D materials that 

\begin{figure}[H]
    \centering
\includegraphics[width=1.0\textwidth]{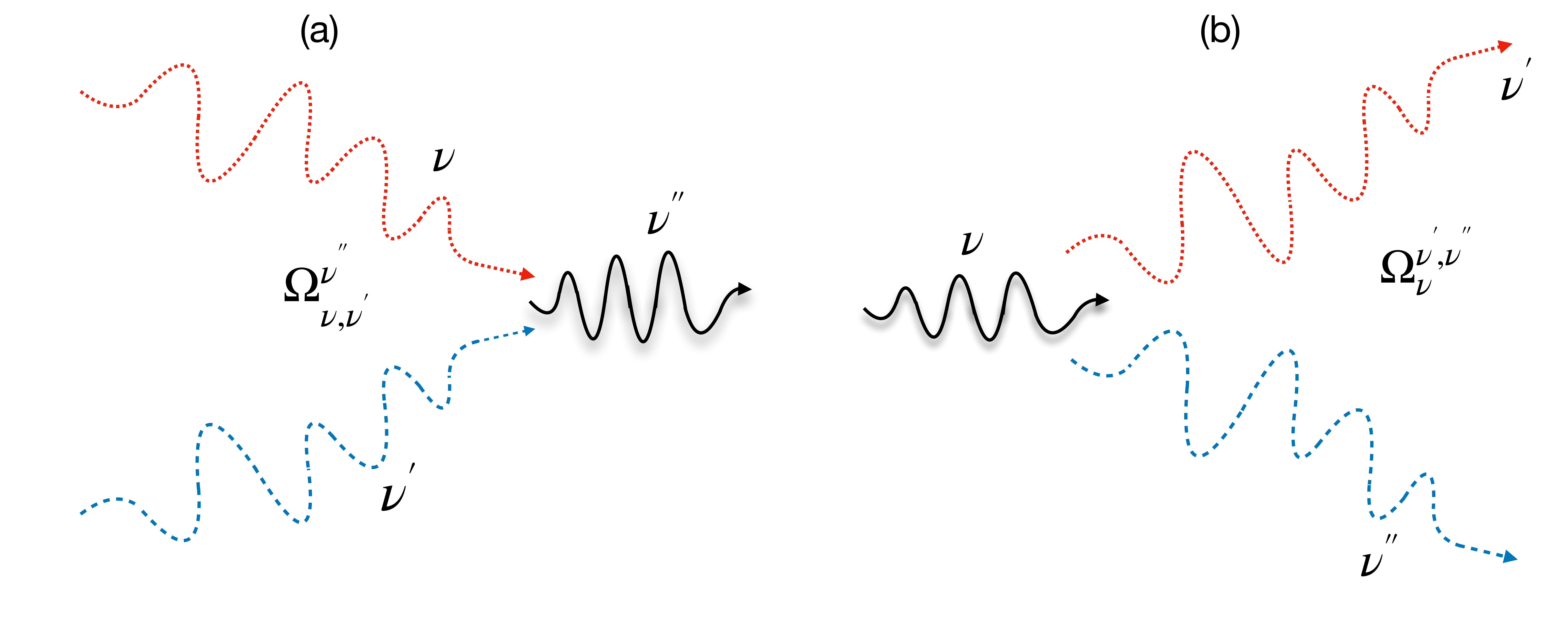}
    \caption{Schematic representations for phonon-phonon scattering: (a) Coalescence and (b) Decay processes. $\Omega_{\nu, \nu^{'}}^{\nu^{''}}$ and $\Omega_{\nu}^{\nu^{'}, \nu^{''}}$ represent the scattering rates due to coalescence and decay processes respectively. In the coalescence process phonon mode $\nu$ absorbs another phonon mode $\nu^{'}$ to yield a third phonon mode $\nu^{''}$. In the decay process phonon mode $\nu$ decays into two phonon modes $\nu^{'}$ and $\nu^{''}$.}
    \label{fig:3ph_scematic}
\end{figure}

\noindent feature phonon hydrodynamics. The parameters that influence phonon hydrodynamics, have been reviewed and the manipulation of such parameters to tune phonon hydrodynamics for better thermal management, has been studied in section \ref{sec:6} followed by section \ref{sec:7} that demonstrates the summary of the review and a detailed future outlook to the problem.

\section{The emergence of phonon hydrodynamics: Approaching from the phonon-phonon scattering rates}{\label{sec:2}}

Phonon hydrodynamics is a specific situation arising in a crystal lattice when thermal transport in general and phonon transport in specific, happens collectively. This collective motion of phonon quasiparticles resembles fluid flow in fluid hydrodynamics. Now, in phonon transport mechanisms, both harmonic and anharmonic forces are at work. However, it is the anharmonic forces on phonons that drive the phonon scattering and eventually lead to a finite thermal conductivity for any materials. Therefore, persisting coherent motion of phonons, a.k.a phonon hydrodynamics, depends crucially on the three-phonon scattering or the anharmonic phonon-phonon scattering processes.

\subsection{Three-Phonon scattering process}

Anharmonicity in crystal lattice arises from the interaction between lattice vibrational waves, known as phonon-phonon scattering in the quasiparticle picture of solids. Thermal conductivity is one such physical effect which depends heavily on this anharmonic phonon scattering processes. As harmonic phonons do not scatter, anharmonic phonons and their scattering processes are crucial to yield finite thermal conductivity of materials at all temperatures. Figure \ref{fig:3ph_scematic} presents a schematic diagram for three-phonon scattering processes of phonons. Anharmonic phonon scattering can lead to either a coalescence process, where a phonon mode ($\nu$) absorbs another phonon mode ($\nu^{'}$) after scattering and yields a third phonon mode ($\nu^{''}$) or a decay process, where a phonon mode ($\nu$) decays into two phonon modes after scattering ($\nu^{'}$, $\nu^{''}$).

At this juncture of the discussion, we recall the seminal work of Peierls \cite{Peierlsbook1955} which states that anharmonicity alone is not enough to induce thermal resistance in solids. Conservation of momentum in phonon-phonon scattering process leads to the infinite thermal conductivity. Therefore momentum destroying phonon-phonon scattering plays a crucial role in introducing thermal resistance in solids. Thus, three-phonon scattering processes can be divided into two classes: (a) normal scattering (N scattering) and (b) Umklapp scattering (U scattering). N scattering conserves phonon momentum whereas U scattering doesn't. N scattering only redistributes momentum amongst various phonon modes while U scattering gives rise to the thermal resistance. For a typical three-phonon absorption process, as described in Fig \ref{fig:3ph_scematic}.(a), the wave vectors satisfy 
\begin{equation}{\label{eq1}}
    \mathbf{q} + \mathbf{q^{'}} = \mathbf{q^{''}} + \mathbf{G}
\end{equation}

where \textbf{q} and \textbf{q$'$} are the two wave vectors of two scattering phonons with frequencies $\nu$ and $\nu^{'}$ respectively (Fig \ref{fig:3ph_scematic}.(a)), \textbf{q$''$} is the wave vector of phonon created in the absorption process (with frequency $\nu^{''}$ in Fig \ref{fig:3ph_scematic}.(a)) and \textbf{G} represents the reciprocal lattice vector. If \textbf{G} = 0, then the scattering is momentum conserving and therefore designated as N scattering whereas any finite, nonzero value of \textbf{G} indicates a momentum destroying U scattering event. In other words, if the resultant wave vector, after a three phonon scattering process, exceeds the first Brillouin zone (the Wigner-Seitz unit cell in reciprocal lattice \cite{Leebookchapter2020}), \textbf{G} is employed to bring the resultant vector back to the first Brillouin zone at the cost of reversal of the 

\begin{figure}[H]
    \centering
\includegraphics[width=1.0\textwidth]{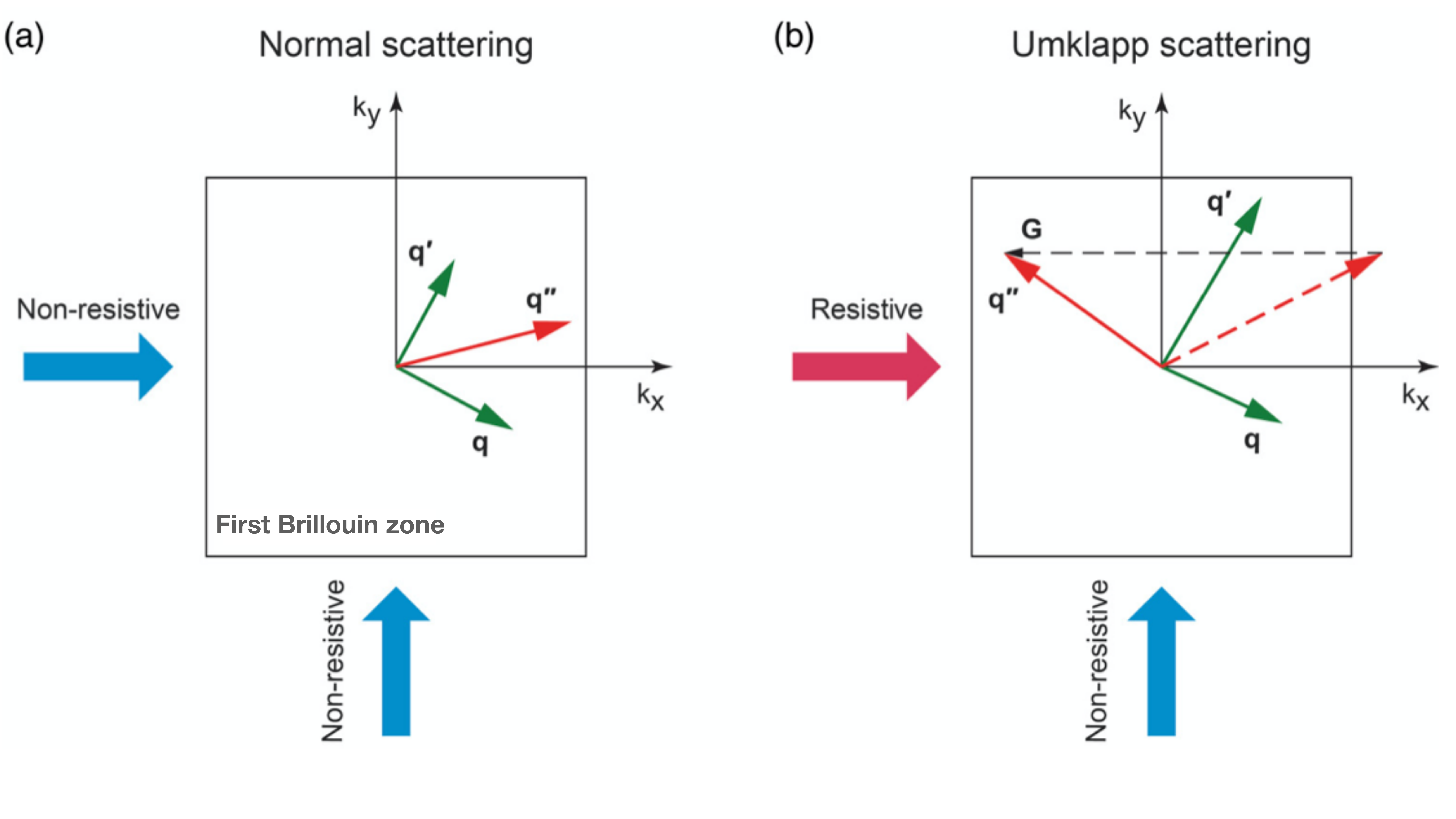}
    \caption{Schematic representation of normal scattering (N scattering) and Umklapp scattering (U scattering) processes. (a) N scattering conserves momentum as the scattering between wave vectors \textbf{q} and \textbf{q$'$} yields \textbf{q$''$} which stays inside the first Brillouin zone of the reciprocal lattice. Also, N scattering does not produce thermal resistance in either the $x$ or the $y$ direction. (b) U scattering between phonon wave vectors \textbf{q} and \textbf{q$'$} yields \textbf{q$''$} which exceeds the first Brillouin zone and therefore destroys the phonon momentum conservation. U scattering is shown to cause resistance in the $x$ direction but not in the $y$ direction. Figure \ref{fig:N_U_schematic} adapted with permission from Ref. \cite{Chen2018R}. Copyrighted by the American Physical Society.}
    \label{fig:N_U_schematic}
\end{figure}

\noindent phonon propagation direction and therefore causing non conservation of quasi momentum of phonons. In this context, we mention a study by Ding $\textit{et al.}$ \cite{Chen2018R} where the concept of N and U scattering processes are redefined. According to this study, a phonon-phonon scattering process is called N scattering if phonon momentum is conserved in the direction of the heat flow and U scattering does not involve in thermal resistance unless the projection of phonon momentum in the scattering process is not conserved in the direction of heat flow. Thus Eq. \ref{eq1} can be modified as \cite{Chen2018R}
\begin{equation}{\label{eq2}}
    \mathbf{q_j} + \mathbf{q_{j}^{'}} = \mathbf{q_{j}^{''}} + \mathbf{G_j}
\end{equation}

\noindent Here $\mathbf{j}$ denotes the direction of heat transport, $\mathbf{q_j}$ and $\mathbf{G_j}$ are the projections of vectors $\mathbf{q}$ and $\mathbf{G}$ along $\mathbf{j}$. Figure \ref{fig:N_U_schematic} shows the pictorial representation of U and N scattering processes. It can be understood from Fig \ref{fig:N_U_schematic}.(b) that the momentum conservation breaks only in the $x$ direction while along $y$ direction phonon momentum is conserved. Therefore the thermal resistance originates along only $x$ direction and the scattering can be termed as U scattering.

\subsection{The microscopic origin and criteria for Phonon hydrodynamics}

As we mentioned earlier, phonon hydrodynamics is born out of the collective phonon transport in crystal lattice. Following the discussion of the phonon-phonon scattering rates, the situation of collective transport implies that the phonon momentum dissipates over a sufficiently long time such that within an appreciable time-window the phonons transport in a coherent motion. In the scattering rate perspective, this indicates a situation when N scattering outweighs dissipative scattering of phonons (U scattering, isotope scattering and phonon-boundary scattering). Ever since the pioneering work by Peierls \cite{Peierls1929} on the thermal conduction of phonons in crystal lattice in 1929, the relative importance between momentum conserving and momentum destroying scattering processes were discussed in the community through various analytical works. Here, we introduce a major manifestation of phonon hydrodynamics in a crystal lattice called second sound which deals with the propagation of temperature waves (a detailed discussion is presented in the section 4.1) in a solid. Drawing the ideas from two-fluid theory of He II by Tisza \cite{Tisza1938} and Landau \cite{Landau1941, Landau1947}, Peshkov \cite{Peshkov1944, Peshkov1948} first detected this temperature waves and later Ward and Wilks \cite{WardWilks1, WardWilks2} derived it for an interacting phonon systems with the conservation of collisions. The favourable condition for the occurrence of the second sound had been mentioned in the works by Sussman and Thellung \cite{Sussmann1963} and by Gurzhi \cite{Gurzhi1964} in his investigation on the thermal conductivity of dielectrics at low temperatures, where U scattering events were almost absent. Gurzhi \cite{Gurzhi1964} mentioned the following inequality to hold in a sufficiently massive and pure sample at low temperature to enable phonon hydrodynamics in the form of temperature waves.
\begin{equation}
    l^{N} \ll d \ll l^U
\end{equation}
where $l^N$ and $l^U$ are the effective mean free paths for N and U scattering respectively. A year back, in 1963, Chester \cite{Chester1963} discussed the second sound in solids using a more general form of Fourier heat equation and identified a critical onset frequency $f_c$ = $\frac{1}{2\pi\tau}$ = $\frac{c_s^{2}C}{6\pi \kappa}$, below which thermal wave does not propagate. Here, $\kappa$ is thermal conductivity, $C$ denotes heat capacity per unit volume and $c_s$ is the sound velocity. Soon after, this novel feature in heat transport had experimentally been found in He IV crystals by Mezhov-Deglin \cite{Mezhov-Deglin1965}. 

In one of their series of seminal works, Guyer and Krumhansl \cite{GK1:1966} solved the linearized Boltzmann transport equation (LBTE) for pure phonon field in terms of the eigenvectors of the N process collisional operator and understood the interplay between N and R (resistive) processes in dictating the limiting behavior of thermal conductivity. Solving the LBTE, in their subsequent work \cite{GK2:1966}, they developed a set of macroscopic equations and solved the steady state problem for low temperature phonon gas in one dimensional flow in a cylinder. The existence of another distinct phonon hydrodynamic feature called phonon Poiseuille flow (will be discussed later in detail) was observed which is consistent with the earlier investigation by Sussmann and Thellung \cite{Sussmann1963}. The effect of different phonon scattering events can be understood using the average scattering rates, defined by:
\begin{equation}
    \langle \tau_{i}^{-1} \rangle_{ave} = \frac{\sum_{\lambda}C_{\lambda}\tau_{\lambda i}^{-1}}{\sum_{\lambda}C_{\lambda}}
\end{equation}
Here, $\lambda$ defines phonon modes ($\textbf{q}$, $j$) comprising wave vector $\textbf{q}$ and phonon branch $j$. Index $i$ denotes normal, Umklapp, isotope and boundary scattering processes, denoted by N, U and I and B respectively. $C_\lambda$ is the modal heat capacity, given by 
\begin{equation}
    C_\lambda = k_{B} \left(\frac{\hbar\omega_{\lambda}}{k_{B}T}\right)^{2} \frac{exp(\hbar\omega_{\lambda}/k_{B}T)}{[exp(\hbar\omega_{\lambda}/k_{B}T) -1]^2}
\end{equation}
where, $T$ denotes temperature, $\hbar$ is the reduced Planck constant and $k_B$ is the Boltzmann constant. According to the condition prescribed by Guyer and Krumhansl \cite{GK1:1966, GK2:1966} hydrodynamic regime exists if
\begin{equation}{\label{eq:hydro}}
   \langle \tau_{U}^{-1} \rangle_{ave} \ll \langle \tau_{N}^{-1} \rangle_{ave}
\end{equation}
Moreover, Guyer's condition \cite{GK2:1966} for the occurrence of second sound and Poiseuille's flow reads: 
\begin{equation}{\label{eq:poiseuille}} 
   \langle \tau_{U}^{-1} \rangle_{ave} \ll \langle \tau_{B}^{-1} \rangle_{ave} \ll \langle\tau_{N}^{-1} \rangle_{ave}
\end{equation}

\begin{figure}[H]
    \centering
\includegraphics[width=1.0\textwidth]{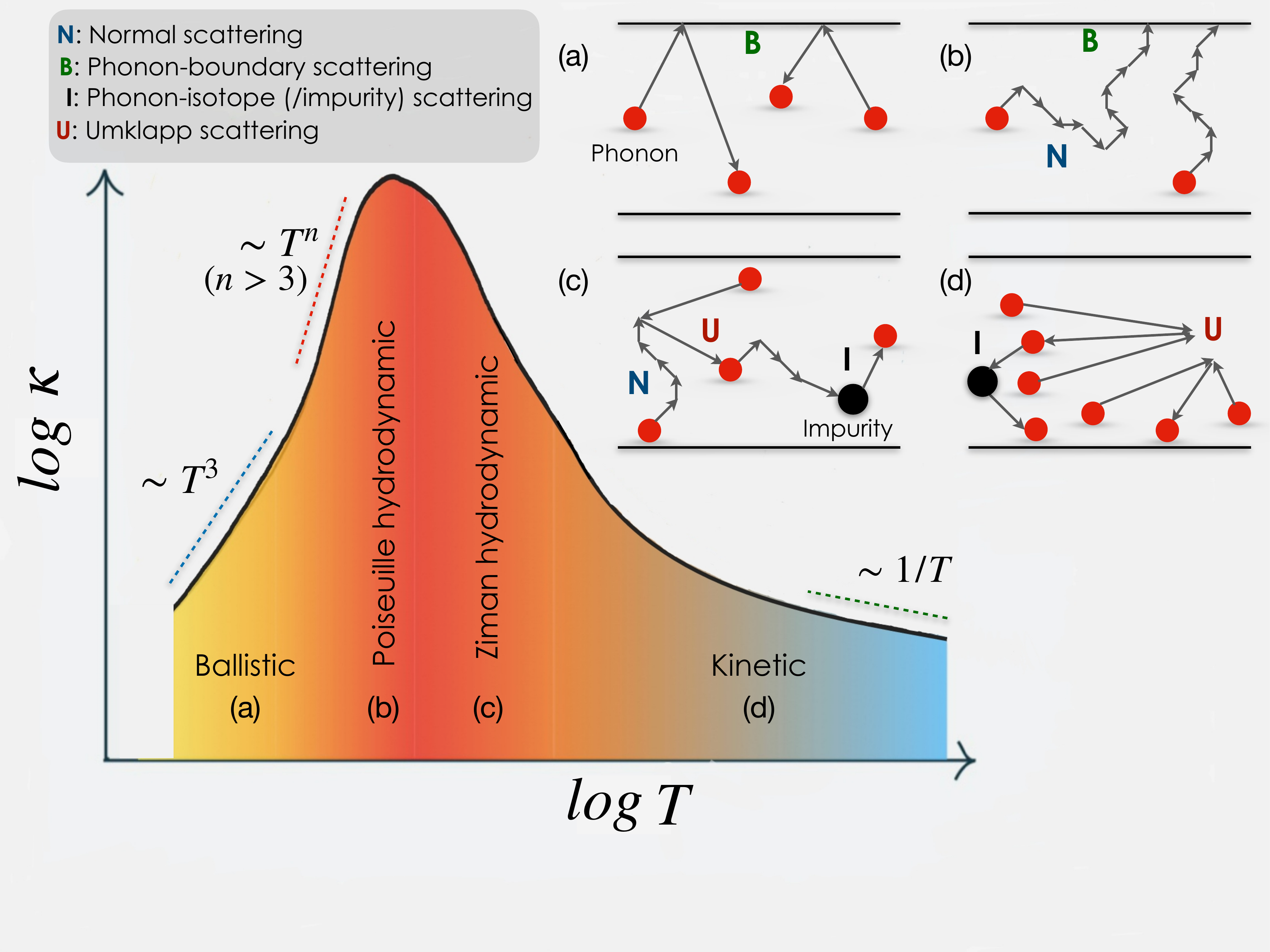}
    \caption{Thermal transport regimes for a generic three dimensional solid is presented through a schematic diagram of the temperature ($T$) variation of the lattice thermal conductivity ($\kappa$). Four distinct thermal transport regimes are shown: (a) ballistic, (b) Poiseuille hydrodynamic, (c) Ziman hydrodynamic and (d) kinetic. In the ballistic regime, $\kappa$ varies as $T^3$ due to purely the effect of specific heat, as phonon mean free path is controlled by the sample size. In the Poiseuille regime, the exponent exceeds $3$. At high temperature limit, $\kappa$ varies as $1/T$ as described by Slack model \cite{slack1964, Morelli2006}. Insets: For each of these regimes, different phonon scattering hierarchies are schematically represented via normal (N), phonon-boundary (B), phonon-isotope (I) and Umklapp (U) scattering events. (a) phonons are directly scattered via boundary scattering. (b) phonons perform N scattering which dissipates via B scattering. (c) phonons perform N scattering which dissipates via R (resistive:U and I) scattering. (d) phonons mostly perform momentum destroying U scattering. We note that for each of these four regimes, shown in insets, only dominant scattering events are illustrated for clarity.}
    \label{fig:regimes}
\end{figure}

\begin{table}[H]
\caption{\label{table1}Thermal transport regimes in terms of average phonon scattering rates ($\langle \tau_{i}^{-1} \rangle_{ave}$) as prescribed by GK conditions \cite{GK2:1966}. Here $i$ denotes N, U, R and B which represent normal, Umklapp, total resistive (Umklapp+isotope) and phonon-boundary scattering respectively.}
\footnotesize
\begin{ruledtabular}
\begin{tabular}{ccc}

Thermal transport regime & Condition(s) & Description\\
\colrule
(A) Ballistic & $\langle \tau_{R}^{-1} \rangle_{ave} \ll \langle \tau_{B}^{-1} \rangle_{ave}$, $\langle \tau_{N}^{-1} \rangle_{ave} \ll \langle \tau_{B}^{-1} \rangle_{ave}$ & Phonon-boundary scattering is dominant\\
& & due to the finite size of the sample.\\
(B) Hydrodynamic & $\langle \tau_{R}^{-1} \rangle_{ave} \ll \langle \tau_{N}^{-1} \rangle_{ave}$, $\langle \tau_{B}^{-1} \rangle_{ave} \ll \langle \tau_{N}^{-1} \rangle_{ave}$ & N scattering dominates over \\
& & B and R scattering.\\
\\
(i) Poiseuille hydrodynamic & $\langle \tau_{R}^{-1} \rangle_{ave} \ll \langle \tau_{B}^{-1} \rangle_{ave} \ll \langle\tau_{N}^{-1} \rangle_{ave}$ & Heat flux is dissipated dominantly via \\
& & extrinsic boundary scattering.\\

(ii) Ziman hydrodynamic & $\langle \tau_{B}^{-1} \rangle_{ave} \ll \langle \tau_{R}^{-1} \rangle_{ave} \ll \langle\tau_{N}^{-1} \rangle_{ave}$ & Heat flux is dissipated dominantly via \\
& & Umklapp and isotope scattering.\\
\\
(C) Kinetic regime &  $\langle \tau_{N}^{-1} \rangle_{ave} \ll \langle \tau_{R}^{-1} \rangle_{ave}$, $\langle \tau_{B}^{-1} \rangle_{ave} \ll \langle \tau_{R}^{-1} \rangle_{ave}$ & Umklapp and isotope scattering  \\
& & events dominate over N and B scattering. \\

\end{tabular}
\end{ruledtabular}
\end{table}

\noindent The advent of highly efficient computational resources over the years as well as several phenomenological models of heat transfer (Callaway \cite{Callaway1959}, Callaway-Holland \cite{Callaway1959, Holland1963}, Klemens \cite{KLEMENS19581}, Slack \cite{slack1964, Morelli2006}) enable accessing the phonon scattering rates corresponding to the N and R processes and therefore Guyer and Krumhansl conditions permit a feasible and robust way to identify the presence of phonon hydrodynamics in various nonmetallic systems and have been used extensively in current state-of-the-art research on phonon physics \cite{Cepellotti2015, Broido, Bitheory2018, Ding2018, kanka2, ZhangPolymer2020}. Figure \ref{fig:regimes} and Table \ref{table1} summarize the Guyer and Krumhansl (GK) conditions for the occurrence of phonon hydrodynamics and demonstrate distinct thermal transport regimes from the temperature variation of lattice thermal conductivity ($\kappa$) of a generic three dimensional material with their specific scattering protocols. Gurevich and Shklovskii \cite{Gurevich1967} also gave similar arguments around the same time on the conditions for realizing the second sound in semiconductors, produced by long-wave phonons in a frequency interval. Hardy \cite{Hardy1970} solved the complete linearized Boltzmann equation in terms of the eigenvectors of collision matrix including normal, Umklapp and isotope scattering processes. He discussed the existence of `driftless' and `drifting' second sound in the hydrodynamic transport regimes of solids \cite{Hardy1970}. It was found that the dominance of N-scattering events are necessary to feature `drifting' second sound while  a uniform energy flux with an exponential decay is essential to feature the `driftless' second sound. Hardy's analysis \cite{Hardy1970} was stressed upon the fact that the slow decay of energy flux is the most essential criteria for phonon hydrodynamics. The analysis \cite{Hardy1970} also mentioned that the domination of N scattering is not always necessary for the existence of second sound, but when it dominates, phonon hydrodynamics seems to be observable.

All the studies discussed above, are aligned with the same idea of the identification of a window in the relaxation time spectrum that supports hydrodynamic features of the phonon gas. At this point, for the clarity of the readers, we tend to briefly discuss the hydrodynamic conditions, in terms of scattering rates, that had been discussed in the study of Beck $\textit{et al.}$ \cite{BeckReview1974}. A system of phonon gas can be described by a distribution function $f(q,\textbf{r},t)$ and its time evolution is dictated by the Peierls-Boltzmann equation (a detailed account is given in Section 4.1.2). The energy is conserved during the collisions between phonons. However, the phonon hydrodynamics demands a dominance of N scattering over resistive scattering events and therefore dictates the conservation of quasimomentum of the phonons throughout the crystal. This situation invokes a drift to the phonon distribution function in thermal equilibrium given by the displaced distribution ($f_{BE}^{d}$)
\begin{equation}{\label{eq:drift}}
    f_{BE}^{d} = \frac{1}{exp[\beta\hbar(\omega-\mathbf{q}\cdot \mathbf{u})]-1}
\end{equation}
where $\beta$ = $1/k_{B}T$, $\textbf{u}$ is the drift velocity of the phonon gas and $\textbf{q}$ is the phonon wave vector. It is noted that both $\textbf{u}$ and $\beta$ are space and time-dependent corresponding to local thermal equilibrium. Using Peierls-Boltzmann equation, two conservation laws corresponding to energy and momentum, involving partial derivatives of time and position yield respectively
\begin{equation}{\label{eq:energy2}}
    \frac{\partial}{\partial t} E\left(\mathbf{r},t\right) + \frac{\partial}{\partial r_{i}} Q_{i}\left(\mathbf{r},t\right) = 0
\end{equation}

\begin{equation}{\label{eq:momentum2}}
    \frac{\partial}{\partial t} P_{i}\left(\mathbf{r},t\right) + \frac{\partial}{\partial r_{j}} P_{ij}\left(\mathbf{r},t\right) = 0
\end{equation}
where $E(\textbf{r},t)$, $Q_{i}(\textbf{r},t)$, $P_{i}(\textbf{r},t)$, and $P_{ij}(\textbf{r},t)$ designate densities of energy, energy current along a specific direction $i$ ($x$, $y$, or $z$), momentum along $i$, and momentum flux along $j$ respectively. Eq. \ref{eq:energy2} and \ref{eq:momentum2} represent hydrodynamic equations as long as $E(\textbf{r},t)$, $Q_{i}(\textbf{r},t)$, $P_{i}(\textbf{r},t)$, and $P_{ij}(\textbf{r},t)$ can be expressed in terms of the hydrodynamic variables $\beta(\textbf{r},t)$ and $\textbf{u}(\textbf{r},t)$. Approximating Debye model with $\omega_{\textbf{q}} = c_{\lambda}\mid \mathbf{q}\mid$ and employing mean free time approximation used by Sussmann and Thellung \cite{Sussmann1963}, the conservation laws Eq. \ref{eq:energy2} and \ref{eq:momentum2}  respectively become

\begin{equation}{\label{eq:energynew1}}
    \frac{\dot{\beta}}{\beta_{0}} = \frac{1}{3} \nabla \cdot \mathbf{u} + \frac{1}{3} \frac{\nabla^{2}\beta}{4\epsilon\beta_{0}}\sum_{\lambda}\sigma_{\lambda}(c_{\lambda}^{2}-3c_{II}^{2})
\end{equation}

\begin{equation}{\label{eq:momentumnew1}}
     \resizebox{.78\hsize}{!}{%
     $\dot{\mathbf{u}_{i}} = 3c_{II}^{2} \frac{\nabla_{i}\beta}{\beta_{0}}-3c_{II}^{2}\frac{\nabla_{i}\beta}{4\epsilon\beta_{0}}\sum_{\lambda}\sigma_{\lambda}(1-\frac{3c_{II}^{2}}{c_{\lambda}^{2}})+\frac{3c_{II}^{2}}{20\epsilon}(\frac{1}{3}\nabla_{i}\nabla \cdot \mathbf{u} + \nabla^{2}u_{i})\sum_{\lambda}\sigma_{\lambda}$%
     }
\end{equation}
Here, $\lambda$ stands for phonon polarization, $c_{\lambda}$ is speed of sound, $\beta_{0}$ is the equilibrium value for inverse temperature, $\epsilon$ is mean thermal energy density, $\sigma_{\lambda}$ $\equiv$ $c_{\lambda}^{2}$ $\sum_{\mathbf{q}}q^{2}m(\omega_{q})\tau_{q}$, $\tau_{q}$ is the isotropic relaxation time in the mean free time approximation, and $c_{II}$ denotes the speed of second sound \cite{BeckReview1974}. Eliminating $\mathbf{u}$ from Eq. \ref{eq:energynew1} and \ref{eq:momentumnew1} gives rise to equation for damped second sound
\begin{equation}{\label{eq:damped}}
    \ddot{\beta} - c_{II}^{2}\nabla^{2}\beta - 2c_{II}^{2}\tau_{II}\nabla^{2}\dot{\beta} + \frac{\dot{\beta}}{\tau_{R}} = 0
\end{equation}
where $\tau_{II}$ and $\tau_{R}$ are the relaxation times corresponding to Normal and resistive scattering respectively. The ansatz
\begin{equation}
    \beta(r,t) - \beta_{0} = \int d\mathbf{q} e^{i(\mathbf{q}\cdot \mathbf{r}-\Omega t)}\beta(\mathbf{q}, \Omega(q))
\end{equation}
gives rise to
\begin{equation}
    \Omega^{2} + \frac{i\Omega}{\tau_R} + 2i\Omega\tau_{II}c_{II}^{2}q^{2} = c_{II}^{2}q^{2}
\end{equation}
which eventually leads to the dispersion relation
\begin{equation}
    \Omega = \pm c_{II}q \sqrt{1-\frac{1}{c_{II}^{2}q^{2}}\left(\tau_{II}c_{II}^{2}q^{2} + \frac{1}{2\tau_{R}}\right)^{2}} - i\left(\tau_{II}c_{II}^{2}q^{2} + \frac{1}{2\tau_{R}}\right)
\end{equation} 
If the damping of the second sound has to be small, the period of temperature perturbation should simultaneously follow $\Omega \tau_{II}$ $\ll$ 1 (abundance of Normal scattering) and $\Omega \tau_{R}$ $\gg$ 1 (rare resistive scattering), leading to the following condition
\begin{equation}
    \tau_{R}^{-1} \ll \Omega \ll \tau_{II}^{-1}
\end{equation}

\section{The emergence of phonon hydrodynamics: Approaching from the collective excitation perspective}{\label{sec:3}}

The backbone of the theories of phonon thermal transport lies in solving the Boltzmann transport equation (BTE) for phonons or using Green-Kubo approach in the realm of linear-response theory. The analysis of phonon scattering rates seems to be a feasible approach to investigate phonon hydrodynamics within the relaxation time approximation of the phonon gas. However, in the advent of powerful computational resources and related development in the field of computations with various ab-initio accurate techniques, several methods have been discovered to directly solve BTE without simplifications and assumptions \cite{chaput}. Along this line of thought, in an alternative approach to understand the microscopic origin of the collective phonon dynamics, the failure of the single-mode relaxation time approximation (SMA or RTA) seems to be a key to detect collective phonon transport \cite{Lindsay2019}. 

For several materials, experimentally observed thermal conductivity had often been reproduced in a surprisingly accurate manner using BTE beyond the single mode relaxation time approximation (SMA or RTA) \cite{chaput, Cepellotti2015, Bitheory2018}. However, the departure from RTA approach of phonon gas costs the conceptual complications \cite{Lindsay2019}. This is due to the fact that the full solution of BTE abandons the approach of phonon relaxation times and they are no longer relevant descriptors \cite{chaput,Lindsay2019}. Cepellotti and Marzari \cite{Cepellotti2016, Andrea2017, Transport_waves_as_crystal_excitations} posed an important question on this regard whether any form of relaxation times can be included in the picture of the full solution of BTE. The single mode relaxation time approximation (SMA or RTA) takes only the diagonal terms of the scattering matrix \cite{Fugallo_2018} into account in solving the LBTE and the closed form solution looks like
\begin{equation}
    \frac{1}{V} \sum_{\lambda'} \Omega_{\lambda \lambda'} \Delta f_{\lambda'}\left(\mathbf{x}, t\right) \approx \frac{\Delta f_{\lambda}\left(\mathbf{x}, t\right)}{\tau_{\lambda}^{RTA}}
\end{equation}
where $V$ is the normalization volume, $\lambda$ $\equiv$ ($\textbf{q}$, $j$) denotes the phonon modes with specific wave vector ($\textbf{q}$) and phonon branch $j$. $\Omega_{\lambda \lambda'}$ is the linear phonon scattering operator and $\Delta f_{\lambda}$ = $f_{\lambda}$ - $\overline{f}_{\lambda}$ stands for the deviation of the phonon distribution from equilibrium Bose-Einstein distribution $\overline{f}_{\lambda} (\mathbf{x}, t)$ = $\left[exp(\hbar \omega_{\lambda}/k_{B}T)-1 \right]^{-1}$. The approach of collective phonon excitation, on the other hand, takes the whole scattering matrix into account as well as the direct solution of LBTE without any approximation. In this picture, Cepellotti and Marzari showed \cite{Cepellotti2016} that the collective excitations can be written as an eigenvalue equation 

\begin{figure}[H]
    \centering
\includegraphics[width=1.0\textwidth]{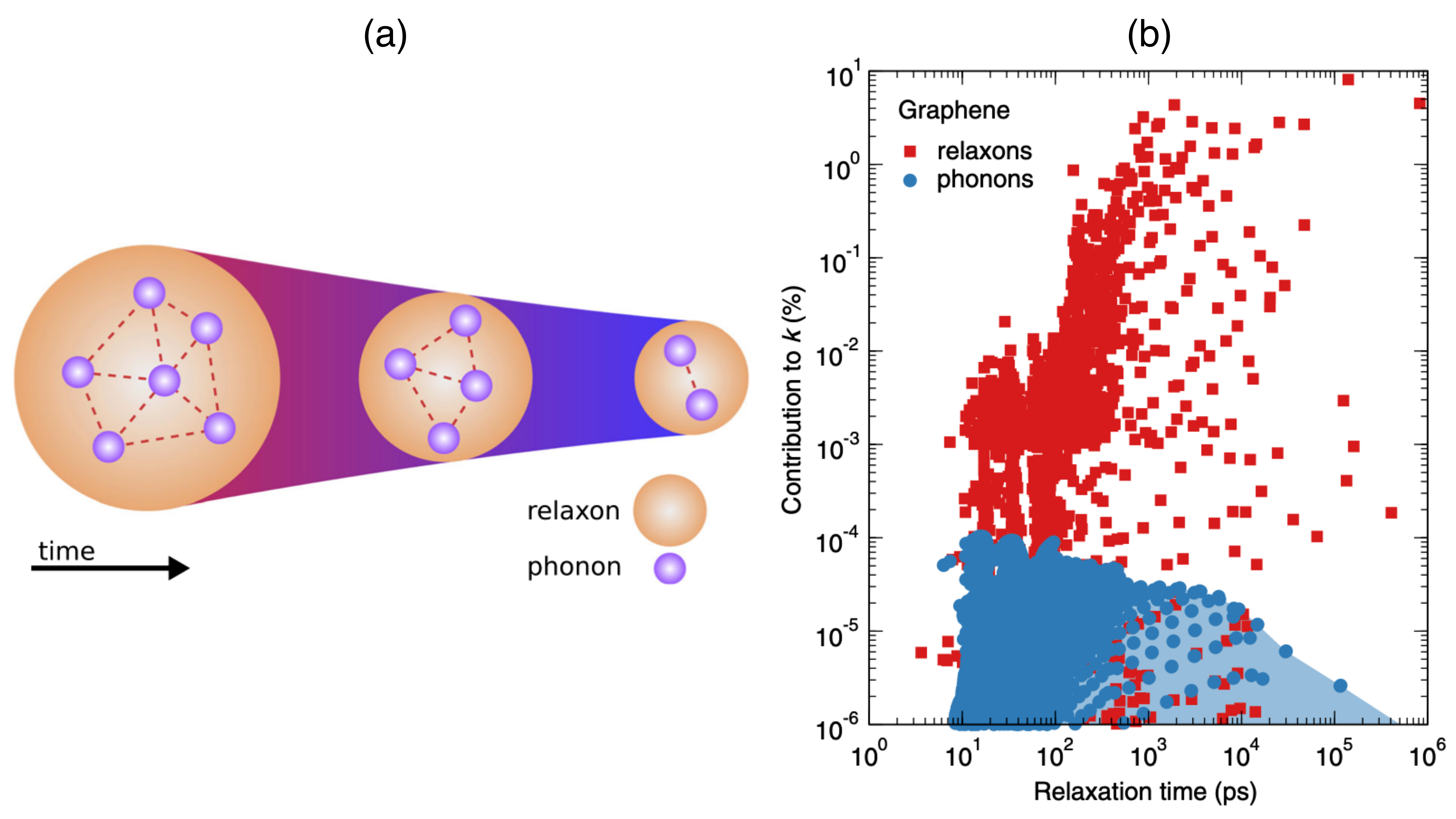}
    \caption{(a) Schematic presentation of the `relaxon'. Each of the relaxons consists of a linear combination of phonons that scatter within themselves but are decoupled from phonons belonging to different relaxons. Reprinted with permission from Ref. \cite{Cepellotti2016}. \href{https://creativecommons.org/licenses/by/3.0/}{CC BY 3.0}. (b) The contribution of relaxation times (considering both the heat carriers: phonons and relaxons) to the thermal conductivity of graphene at room temperature is presented. Relaxons are shown to possess longer lifetimes than that of the single phonon excitations. A significant contribution of relaxons to the thermal conductivity is observed at relaxation times greater than $10^3$ ps while phonons contribute to thermal conductivity mostly in the range between 10 to 100 ps. Phonons are realized as a continuous spectrum while relaxons are discrete and smaller number of relaxons are found to be sufficient (as the contribution to the thermal conductivity is significantly higher than that of the phonons) to accurately represent the thermal conductivity of graphene. Reprinted with permission from Ref. \cite{Cepellotti2016}. \href{https://creativecommons.org/licenses/by/3.0/}{CC BY 3.0}.}
    \label{fig:relaxon1}
\end{figure}

of the form 
\begin{equation}{\label{eq:relaxon}}
    \frac{1}{V} \sum_{\lambda'} \tilde{\Omega}_{\lambda \lambda'} \theta_{\lambda'}^{\alpha} = \frac{1}{\tau_{\alpha}}\theta_{\lambda}^{\alpha}
\end{equation}
where $\tilde{\Omega}_{\lambda \lambda'}$ is the scaled, real symmetric scattering matrix, defined as 
\begin{equation}
    \tilde{\Omega}_{\lambda \lambda'} = \Omega_{\lambda \lambda'} \sqrt{\frac{\overline{f}_{\lambda'}(\overline{f}_{\lambda'}+1)}{\overline{f}_{\lambda}(\overline{f}_{\lambda}+1)}}
    \end{equation}
As $\tilde{\Omega}$ is real, symmetric matrix, it can be diagonalized with eigenvectors $\theta_{\lambda'}^{\alpha}$ and real eigenvalues 1/$\tau_{\alpha}$ such that it satisfies Eq. \ref{eq:relaxon}. Thus, it was shown \cite{Cepellotti2016} that collective phonon excitations can be expressed in terms of a characteristic relaxation time ($\tau_{\alpha}$). However, the relaxation time corresponds to a collective excitation or `relaxon' instead of single excitation of phonons as realised using RTA approach. Each relaxon describes a distribution of out-of-equilibrium or disturbed phonon wave packets and therefore emerges as a linear combination of phonons, which scatter among themselves but decoupled to the phonons that belong to different relaxons \cite{Andrea2017, Cepellotti2016}. This enables the use of kinetic theory on the relaxon gas to interpret thermal conductivity in materials.  

\noindent Figure \ref{fig:relaxon1}.(a) presents the schematic representation of relaxon, taken from the work of Cepellotti and Marzari \cite{Cepellotti2016}. Figure \ref{fig:relaxon1}.(b) shows a specific example of thermal conductivity of graphene at room temperature, comparing the relative importance of the relaxation times of the relaxons and that of the phonons, obtained from the RTA approach \cite{Cepellotti2016}. From this relaxation time spectrum of both relaxon and phonons, relaxons are found to be longer lived than phonons and discretized unlike the continuous spectrum of phonons. The discrete nature (in the region of large values mostly) and at least 2 orders of magnitude larger relaxation times than phonons help the relaxon picture to accurately describe the experimentally observed thermal conductivity of graphene even using a small number of relaxons \cite{Cepellotti2016}. 

\subsection{The conditions for phonon hydrodynamics in relaxon approach}

The condition that leads to the phonon hydrodynamics in the relaxon picture, is essentially based on the departure of heat flux equation from that of the Fourier's law. This leads to the condition to distinguish the diffusive from the hydrodynamic transport regime using viscous heat equations. Simoncelli and co-authors investigated \cite{Simoncelli2020} this hydrodynamic deviations from Fourier's law depending on sample's size and reference temperature $\overline{T}$. Novel viscous heat equations were solved \cite{Simoncelli2020} for different sample sizes and different reference temperatures for graphite, diamond and silicon for comparison. The normalized difference ($\mathcal{L}^2$) between the predicted temperature profile by the viscous heat equations and the Fourier's law for a given sample length $l_{tot}$ and reference temperature $\overline{T}$, serves as one such parameter that can be computed numerically. It is defined \cite{Simoncelli2020} as

\begin{figure}[H]
    \centering
\includegraphics[width=1.0\textwidth]{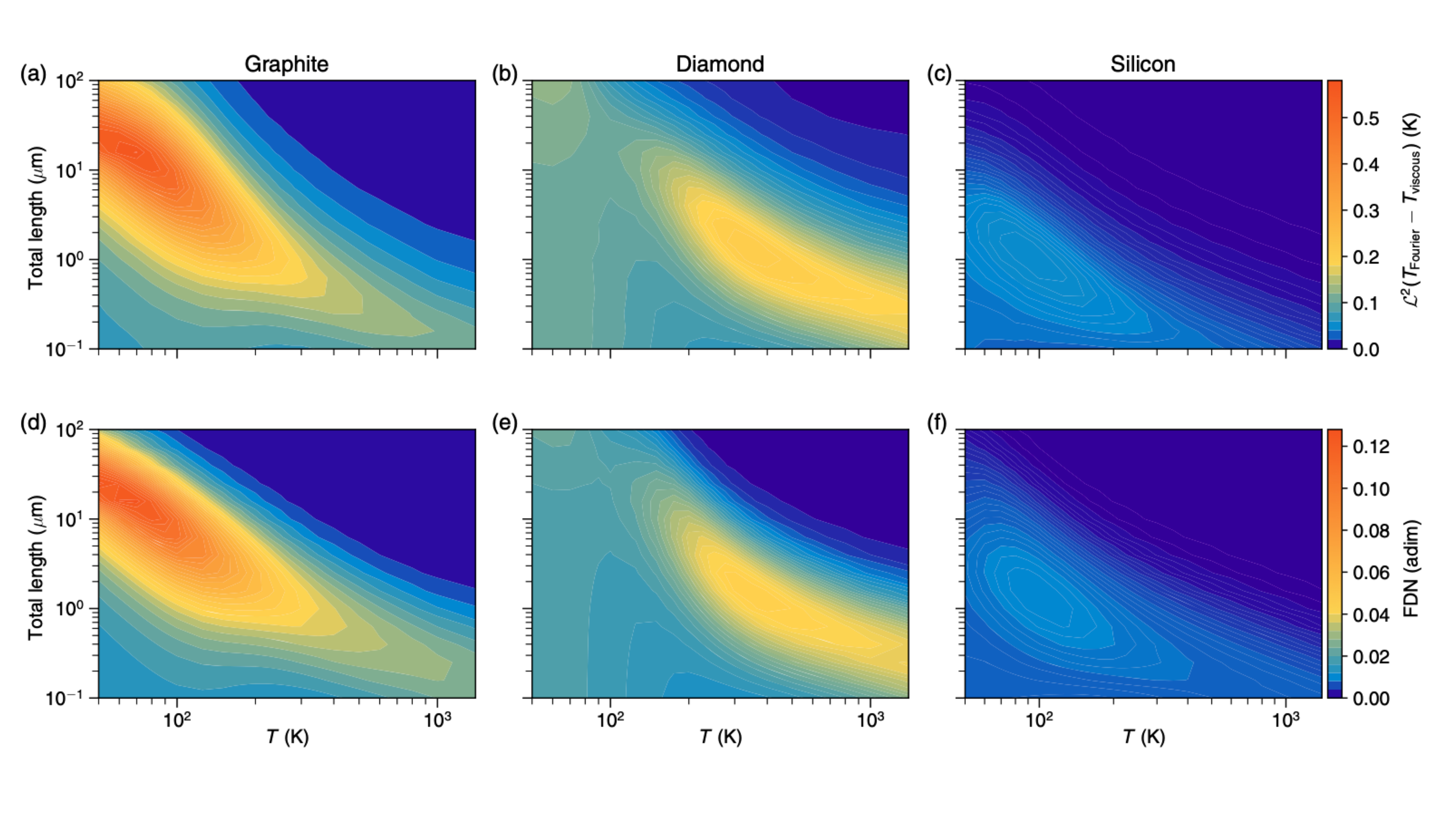}
    \caption{Top row: Temperature variation of $\mathcal{L}^{2}$ distances between the temperature profile predicted by the Fourier’s law and that of the viscous heat equations derived by Simoncelli $\textit{et al.}$ \cite{Simoncelli2020}, are shown as a function of total length of a sample of (a) graphite, (b) diamond, and (c) silicon. The color code quantifies the $\mathcal{L}^{2}$ which is a descriptor to measure phonon hydrodynamic effects via the deviation parameter. Bottom row: Fourier deviation numbers (FDN) are presented for the same materials as a function of temperature and total length of the sample. The FDN description is found to be consistent with the $\mathcal{L}^{2}$ description in identifying the hydrodynamic regimes in terms of temperature and characteristic length. Both the top and the bottom rows represent a strong hydrodynamic effect in graphite even up to 100 K, while diamond and silicon show a comparatively mild and almost no signature of phonon hydrodynamics respectively. Reprinted with permission from Ref. \cite{Simoncelli2020}. \href{https://creativecommons.org/licenses/by/4.0/}{CC BY 4.0}.}
    \label{fig:relaxon3}
\end{figure}

\begin{equation}
    \mathcal{L}^{2} \left[T_{Fourier} - T_{viscous}\right] \left(l_{tot}, \overline{T}\right) = \sqrt{\frac{\int_{G} \left[T_{Fourier} (x, y) - T_{viscous} (x, y)\right]^{2}dxdy}{\int_{G} dxdy}}    
\end{equation}
where $G$ is defined by the integration variable $x$ as $x$ $>$ $\frac{1}{5} l_{tot}$, corresponding to the spatially homogeneous region of the sample. However, to capture the essence of this deviation in a computationally cheaper way, rewriting viscous heat equations in reduced units, Simoncelli $\textit{et al.}$ \cite{Simoncelli2020} defined a quantity called Fourier deviation number (FDN), described as
\begin{equation}
    FDN = \left(\frac{1}{\pi_{1}} + \frac{1}{\pi_{3}} \right)^{-1}
\end{equation}

\noindent where dimensionless parameters $\pi_{1}$ = $\frac{\sqrt{\overline{T}AC}Wu_{0}L}{\kappa \delta T}$ and $\pi_{3}$ = $\frac{\mu}{D_{U}L^{2}A}$. Here $C$ is specific heat, $\mu$ is the thermal viscosity and $A$ and $W$ are two parameters related to $\mu$. $A$ is specific momentum, defined as $A = \partial P/\partial u$ = $\frac{1}{k_{B}\overline{T}V}\sum_{\lambda} \overline{f}_{\lambda}(\overline{f}_{\lambda}+1)(\hbar q)^{2}$, where $u$ is drift velocity, and $W$ is the velocity of relaxon, obtained by projecting phonon group velocity into the momentum conserving eigenvector as $W_{\alpha}$ = $\frac{1}{V}\sum_{\lambda}\phi_{\lambda}v_{\lambda}\theta_{\lambda}^{\alpha}$. $L$ is characteristic size, $u_0$ is the drift velocity, $\delta T$ is the temperature perturbation and $D_{U}$ is the momentum dissipation rate. These notations can also be extended to the 3D case with full tensorial notations as mentioned in \cite{Simoncelli2020}.

\noindent Two conditions are needed to visualize appreciable hydrodynamic feature in this picture: (a) The coupling between drift velocity and temperature needs to be large for the deviation between Fourier's law and viscous heat equations, i.e $\pi_{1}$ $\gg$ 1 and (b) Viscous effects should dominate over crystal-momentum dissipation, i.e $\pi_{3}$ $\gg$ 1. Therefore, large hydrodynamic effect is expected if the following condition is satisfied
\begin{equation}
    FDN = \left(\frac{1}{\pi_{1}}+\frac{1}{\pi_{3}} \right)^{-1} \gg 1
\end{equation}

\noindent Figure \ref{fig:relaxon3}, taken from \cite{Simoncelli2020}, describes the aforementioned difference between Fourier's law and viscous heat equations via $\mathcal{L}^2$ and FDN as descriptors to distinguish phonon hydrodynamics in graphite, diamond and silicon. The strong signature of deviation from the Fourier's law is observed for graphite (Fig \ref{fig:relaxon3}.(a)), which is known to feature hydrodynamics, at low temperature and large sample size limit \cite{Ding2018}. Diamond, having a large thermal conductivity with weak U scattering, is a potential candidate to feature phonon hydrodynamics. In Fig \ref{fig:relaxon3}.(b), the $\mathcal{L}^2$ parameter is seen to predict the largest deviation from Fourier's law in diamond around room temperature and for sample size $>$ 1 $\mu$m. However, compared to the graphite, hydrodynamic phonon signatures seem to be feeble in diamond. For silicon, a very small deviation is observed, mostly at low temperatures as can be seen in Fig \ref{fig:relaxon3}.(c), making it not prone to the phonon hydrodynamic behavior. This is also consistent with earlier studies on silicon as the RTA approximation yields similar values as that of the full LBTE solution for silicon \cite{Sibroido2007, Giorgia2013, Cepellotti2016}. Figure \ref{fig:relaxon3}.(d), (e) and (f) present a consistent picture of the deviation from Fourier's law and the possibilities of displaying hydrodynamic effects for these three materials, captured via FDN. With a negligible computational cost, FDN seems to emerge as a perfect predictor to identify phonon hydrodynamics in materials in the relaxon picture.

\section{Features associated with phonon hydrodynamics: Phenomenological viewpoint}{\label{sec:4}}

The phenomena of phonon hydrodynamics manifest themselves in some peculiar experimentally observable or theoretically realized features that are the representative signatures of the collective motion of phonons. Historically, through these signatures, the presence of phonon hydrodynamics was first introduced within the subject of physics related to phonons. Phonons follow the Bose-Einstein distribution in equilibrium which can be perturbed by a temperature gradient with an abundance of the N scattering events. The N scattering events allow coherent phonon flow and transform the equilibrium phonon distribution to a displaced Bose-Einstein type with a drift velocity associating with it. This indicates gaining of excess momentum to the phonons which shuttles through the N scattering events in such a way that all phonon modes adopt the same drift velocity \cite{Broido}. Umklapp and other resistive scattering events cause non conservation of phonon momentum and force the phonon modes to relax back to the equilibrium Bose-Einstein distribution. Therefore to realize various features of phonon hydrodynamics, the timescale of the phonons to sustain the displaced BE distribution is crucial. In this section, we will address few such prominent signatures of phonon hydrodynamics, the physics behind them and the state-of-the-art account of their research.

\subsection{\textbf{Second sound}} 

In general, `second sound' is referred to the propagation of heat as weakly damped waves in contrary to the usual diffusive propagation of heat in a solid. The propagation of a heat pulse in a solid varies distinctively depending on the relative weight-age of the Normal, Umklapp and other resistive scattering (phonon-boundary, phonon-isotope etc) processes in a solid. In the hydrodynamic transport regime, the heat pulse is carried mostly by very many N scattering events, leading to a propagation of weakly damped (due to very less resistive scattering events) phonon density waves through the solid, which is called second sound. In the diffusive propagation regime of heat, Fourier's law is obeyed and the strong presence of resistive scattering events suppress the collective motion of phonons supported by N scattering. As a result, the heat pulse can't propagate in the solid and the thermal energy in the heat pulse diffuses \cite{Leebookchapter2020, Broido}. In the ballistic heat propagation regime, the heat pulse propagates but the average phonon mean free path is always greater than the sample size. Therefore collective motion of phonons can occur only in the hydrodynamic regime, featuring `second sound' phenomenon. The name `second sound' comes from its phenomenological similarity with acoustic sound, which propagates in fluids as a pressure wave.

\subsubsection{Historical account:} The idea of second sound surfaced with the Laszlo Tisza's idea of two-fluid theory of liquid helium \cite{Tisza1938}. Liquid helium showed very different features below the so-called lambda transition temperature near 2.2 K and was termed as Helium II. Tisza proposed the idea that the Bose-Einstein condensed fraction of helium II \cite{London1938} can form a superfluid which passes through narrow tubes without any dissipation, whereas, the uncondensed atoms behave as normal fluids \cite{Russel2009}. This leads to the idea of `two-fluid' theory in liquid He. Naturally, two-fluid equations of motions not only predicted density fluctuations of the fluid, but also the temperature or entropy fluctuations which later was termed as `second sound' by Lev Landau \cite{Landau1941}. To conceptualise the `two-fluid' theory of He II, around 1947, Landau put forward the idea of two types of quasiparticles, namely phonons and rotons \cite{Landau1947}. These rotons \cite{Marisroton, Marisroton2} have been understood as higher energy excitations than phonons. However, the second sound velocity of He II showed notable discrepancies \cite{Russel2009} between Tisza's and Landau's theoretical  models at low temperature (below 1 K). The very first experiments to detect second sound in heluim II was carried out by V. Peshkov \cite{Peshkov1944} in 1944 using a resonator in a glass tube to study the standing waves of second sound in helium II. Finally, with some more refinements, Peshkov's experiment \cite{Peshkov1948} matched with the prediction by the Landau's theory. At the same time, Fairbank \textit{et al.} \cite{Fairbank1947} proposed a different experiment to generate second sound at a liquid-vapor interface by using the reflection of the normal sound in the vapor from the liquid surface. Later, Ward and Wilks \cite{WardWilks1, WardWilks2} directly derived the velocity of second sound from the interacting phonon gas models where collisions conserve momentum. R.B. Dingle \cite{Dingle1952} and F. London \cite{London1954} also proposed similar ideas in 1952 and 1954 respectively. This opens up the research pathway to investigate second sound in crystalline solids.

\subsubsection{Theoretical investigations:}

The early experiments on the second sound in solids trigger the theoretical physics community to delve deeper into the physics of phonons to understand second sound in solids. Around 1960, a lot of theoretical investigations on the speed, occurrence, dispersion and frequency of operation of second sound in solids surfaced in the field. Peierls work \cite{Peierls1929} on the transport phenomena in interacting phonon systems, derived via the Boltzmann's equation for dynamics of gases paved a founding stone to deal with the occurrence and consequences of second sound theoretically. To understand the development in this direction of research, first we shall demonstrate the theoretical underpinnings of the derivation of the speed of second sound through the hydrodynamic equation which arises from the Peierls-Boltzmann transport equation.

The time evolution of phonon distribution function $f(q, \textbf{r}, t)$, governed by Peierls-Boltzmann equation reads \cite{Peierls1929, Hardy1970, BeckReview1974}
\begin{equation}{\label{eq:BTE}}
    \frac{\partial f}{\partial t} + \mathbf{v_{k}} \cdot \frac{\partial f}{\partial \bf{r}} = \left[ \frac{\partial f}{\partial t}\right]_{C}
\end{equation}
\noindent where $\textbf{v}_{k}$ is the phonon group velocity and $\left[ \frac{\partial f}{\partial t}\right]_{C}$ represents the change of the phonon distribution function due to the phonon scattering events. In phonon hydrodynamics, as discussed earlier, we are looking at the conditions where quasi-momentum destroying Umklapp processes die out and the quasi-momentum is conserved. In that case, the phonon distribution function at equilibrium will contain an additional drift term in the expression given as 
\begin{equation}
    f_{BE}^{d} = \frac{1}{exp[\frac{\hbar}{k_{B}T}(\omega-\mathbf{q}\cdot \mathbf{u})]-1}
\end{equation}
where $\bf{u}$ is the drift velocity of the phonon gas and $\bf{q}$ is the phonon wave vector. As mentioned in \cite{Broido}, assuming energy and crystal momentum along the direction of flow ($x$), energy and crystal momentum balance equations, derived using Peierls Phonon Boltzmann transport equation, read
\begin{equation}{\label{eq:energy_balance}}
    \frac{\partial}{\partial t} \left( \sum_{\lambda} \int \omega f d\mathbf{q} \right) + \frac{\partial}{\partial x} \left( \sum_{\lambda} \int \omega v_{x} f d\mathbf{q} \right) = 0
\end{equation}

\begin{equation}{\label{eq:momentum_balance}}
    \frac{\partial}{\partial t} \left( \sum_{\lambda} \int q_{x} f d\mathbf{q} \right) + \frac{\partial}{\partial x} \left( \sum_{\lambda} \int q_{x} v_{x} f d\mathbf{q} \right) = 0
\end{equation}
 where $\lambda$ stands for phonon polarization. Eq. \ref{eq:energy_balance} is said to be the energy balance equation where energy and energy current along a specific direction $\alpha$ ($x$, $y$ or $z$) are given respectively as
 \begin{equation}
    E \left(\mathbf{r}, t\right) = \sum_{\lambda} \omega_{\lambda} f\left(q, \mathbf{r}, t\right)  
 \end{equation}
 
 \begin{equation}
    Q_{\alpha} \left(\mathbf{r}, t\right) = \sum_{\lambda} \omega_{\lambda} v_{\alpha \lambda} f\left(q, \mathbf{r}, t\right)  
 \end{equation}
 
\noindent Similarly, Eq. \ref{eq:momentum_balance} describes the momentum balance equation where momentum density along $\alpha$ and momentum flux along $\beta$ are given by

\begin{equation}
    P_{\alpha}\left(\mathbf{r}, t\right) = \sum_{\lambda} q_{\alpha \lambda} f\left(q, \mathbf{r}, t\right)  
 \end{equation}
 
 \begin{equation}
    P_{\alpha\beta} \left(\mathbf{r}, t\right) = \sum_{\lambda} q_{\alpha} v_{\lambda\beta} f\left(q, \mathbf{r}, t\right)  
 \end{equation}
We note that the energy and momentum balance equations stem from the two principal considerations: (a) The energy conservation due to scattering and (b) the crystal momentum conservation due to the assumption of low resistive scattering. If a small drift velocity is assumed with $\mathbf{q} \cdot \mathbf{u}$ $\ll$ $\omega$, then the displaced Bose-Einstein distribution ($f_{BE}^{d}$) of phonons can be linearized as \begin{equation}
    f_{BE}^{d} \approx f_{BE}^{0} + \frac{\hbar}{k_{B}T}f_{BE}^{0}\left(f_{BE}^{0}+1\right)q_{x}u_{x}
\end{equation}
Here $f_{BE}^{0}$ stands for the equilibrium BE distribution. Putting the value of $f_{BE}^{d}$ and neglecting higher order terms involving small $u_x$, the energy and momentum balance equations read \cite{Broido}
\begin{equation}
    {\label{eq:energy_balance1}}
    \resizebox{.75\hsize}{!}{%
    $\left( \sum_{\lambda} \int \omega \frac{\partial f_{BE}^{0}}{\partial T} d\mathbf{q} \right)\frac{\partial T}{\partial t} +  \left(\sum_{\lambda} \int \omega v_{x}\frac{\hbar}{k_{B}T^{2}} f_{BE}^{0}\left(f_{BE}^{0} +1\right)q_{x} d\mathbf{q} \right)\frac{\partial u_x}{\partial x} = 0$%
    }
\end{equation}

\begin{equation}{\label{eq:momentum_balance1}}
    \resizebox{.75\hsize}{!}{%
    $\left( \sum_{\lambda} \int q_{x} \frac{\hbar}{k_{B}T^{2}} f_{BE}^{0}\left(f_{BE}^{0} +1\right) q_{x} d\mathbf{q} \right)\frac{\partial u_x}{\partial t} +  \left(\sum_{\lambda} \int q_{x} v_{x} \frac{\partial f_{BE}^{0}}{\partial T} d\mathbf{q} \right)\frac{\partial T}{\partial x} = 0$%
    }
\end{equation}
Time derivative of Eq. \ref{eq:energy_balance1} and spatial derivative of Eq. \ref{eq:momentum_balance1} and little algebraic exercise gives rise to the hyperbolic wave equation for second sound as 
\begin{equation}
    \frac{\partial^{2}T}{\partial t^{2}} = v_{II}^{2}\frac{\partial^{2}T}{\partial x^{2}}
\end{equation}
Here $v_{II}$ denotes the speed of second sound where 

\begin{equation}
    v_{II} = \sqrt{\left(\frac{\left( \sum_{\lambda}\int q_{x} v_{x} \frac{\partial f_{BE}^{0}}{\partial T} d\bf{q} \right)\left(\sum_{\lambda} \int \omega v_{x} f_{BE}^{0}\left(f_{BE}^{0} +1\right)q_{x} d\bf{q} \right)}{\left(\sum_{\lambda} \int \omega \frac{\partial f_{BE}^{0}}{\partial T} d\bf{q} \right)\left(\sum_{\lambda} \int q_{x}  f_{BE}^{0}\left(f_{BE}^{0} +1\right) q_{x} d\bf{q} \right)} \right)}
\end{equation}

\noindent This derivation by Lee $\textit{et al.}$ \cite{Broido} considered arbitrary phonon dispersion to derive the speed of second sound. At earlier times \cite{WardWilks1, WardWilks2}, most of the studies related to the theoretical prediction of second sound involved the major assumption of the phonon spectrum as a Debye model with three branches. This assumption with $\omega_q$ = $v_{I}\mathbf{q}$, where $v_{I}$ is the speed of acoustic sound leads to the relation between first and second sound as  $v_{II}$ = $v_{I}$ /$\sqrt{3}$. 

\noindent In the early 60's, to understand the phenomena of second sound in solids, macroscopic equations were used \cite{Chester1963} with modifications of the Fourier's heat equation. Sussmann and Thellung \cite{Sussmann1963} derived a more generalized version of the hydrodynamic equation of second sound for a cylindrical domain with rough surface. During this time, a series of theoretical studies by Guyer, Krumhansl and Prohofsky \cite{GK1:1966, GK2:1966, GKdispersion1964, Prohofsky1964} marked as pioneering works in this context. Guyer and Krumhansl solved the linearized Boltzmann equation for phonons as an eigenvector problem of the Normal scattering collision operator. By investigating the steady-state phenomena in a phonon gas, their study \cite{GK1:1966, GK2:1966} explicitly expressed the thermal conductivity as a function of wave vector and frequency and therefore generalized the macroscopic heat equation with Fourier's law at one limit ($\tau_N^{-1}$ $\ll$ $\tau_R^{-1}$) and a macroscopic equation similar to Sussmann and Thellung \cite{Sussmann1963} at the other limit ($\tau_N^{-1}$ $\gg$ $\tau_R^{-1}$). Their seminal works also involved the investigation of the dispersion relation of second sound using the Boltzmann equation of phonon gas with a local temperature perturbation \cite{GKdispersion1964}, the damping \cite{Prohofsky1964} and the operational conditions of second sound in nonmetallic solids in terms of upper and lower frequency bounds \cite{GK2:1966} as mentioned in earlier sections. In the context of dispersion and damping on second sound, Kwok's work \cite{Kwok1967} shed light on the second sound velocity in arbitrary directions for anisotropic solids and found that the damping varies quadratically with the frequency of phonons for anisotropic solids. Later, a similar but simpler form of the second sound velocity had been derived by Maris \cite{Maris1981} for anisotropic solids. The attenuation of second sound was also discussed by Weiss \cite{Weiss1981}. Gurevich and Shklovskii \cite{Gurevich1967} showed that the possibility of occurrence of a damped second sound is related to the large and equal electron and hole concentrations in a semiconductor. Hardy \cite{Hardy1970} envisioned to understand second sound by solving the exact solution of Linearized Boltzmann transport equation(LBTE) using eigenvalues and eigenvectors of the collision matrix. Instead of the conditions driven only by the relative weights of Umklapp and normal scattering, his work introduced a more generic condition for the occurrence of second sound. It was shown that the second sound can propagate if the energy flux decays slow enough for sustaining the temperature wave. Hardy also derived the possibility of `drifting' and `driftless' both kinds of second sounds in crystals, which though envisaged theoretically in some other studies \cite{Enz1968, Varshni1972, DavidSingh1982}, is yet to be validated by experiments \cite{Leebookchapter2020}. A lucid description on second sound using the one-particle densities and the local equilibrium density matrices for phonon fields can be found in the work by Enz \cite{Enz1968}. Ruggeri $\textit{et al.}$ \cite{Ruggeri1996} defined a characteristic temperature and studied its effect on the shape change of the propagating second sound waves in solids.

In an alternative approach, several theoretical works \cite{Gotze1967, Cowley1967, Sham1967, Ranninger1969} on the second sound in solids also rely on the Green's functions method. These approaches can be broadly classified as non-equilibrium and equilibrium Green's functions methods. In the former approach, microscopic derivations of transport equations for phonons are obtained \cite{Horie1964, KwokMartin1966, Meier1969, BeckMeier1970} starting from the lattice Hamiltonian, using phonon number density and following the general prescription by Kadanoff and Baym \cite{kadanoff}. The later approach employed equilibrium Green’s functions procedure to investigate second sound phenomena \cite{Sham1967, Gotze1967, KleinWehner1969, Ranninger1969}. This approach broadly based on the idea that if second sound seems to exist in some system, then irrespective of how it had been excited, it should be realized by some equilibrium correlation function (precisely the autocorrelation function of the energy density) of the system \cite{Ranninger1969}. Other theoretical exploration of second sound involved the effect of strong stationary thermal pulse on second sound \cite{Nielsen1969}, the effect of finiteness of the normal scattering rate on the speed of second sound \cite{HardyAlbers1974}, the effect of pulse propagation along temperature gradients \cite{Coleman1988}, studies on second sound velocities in cubic \cite{Varshni1972} and hexagonal crystals \cite{DavidSingh1982}, calculation of the velocity of drifting second sound in NaF \cite{HardyNaF1971}, NaI \cite{JaswalHardyLiFNaI1972} using anisotropy and dispersion of the phonon frequency spectrum etc.

A thorough account of these theoretical development until 1974 was presented in the work of Beck $\textit{et al.}$\cite{BeckReview1974}. A broader version on all kind of heat waves can be found in the work of Joseph and Preziosi \cite{JosephHeatwaves1989}.

\subsubsection{Experimental methods and observations:}
As predicted by theoretical investigations, solid He satisfies the strict frequency criterion to observe second sound. Therefore, Ackerman and co workers carried out heat-pulse experiments to probe second sound in solid He$^4$ \cite{Ackerman1966solidhelium} and He$^3$ \cite{Ackerman1969solidhe3}. Around the same time, light scattering experiments \cite{Griffin1968review} also employed to probe second sound in solids. As mentioned by Lee \textit{et al.} \cite{Leebookchapter2020}, these experiments can be broadly distinguished as two different methods: (a) Heat-pulse experiments and (b) Light scattering experiments. In literature, both the heat-pulse experiments \cite{Ackerman1966solidhelium, Ackerman1969solidhe3, Ackerman1968, Jackson1970NaF, Jackson1971NaF, McNelly1970NaF, Narayanmurti1972Bi, Rogers1971AlkaliHalide, Narayanmurti1975} and the light scattering methods \cite{Griffin1965, Griffin1968review, Guyer1965light, Wehner1972, Pohl1976NaF, Hehlen1995SrTio3, Koreeda2007SrTio3, Koreeda2009SrTio3, Koreeda2010KTaO3} have been proved to be effective to detect second sound in solids. In a standard heat pulse experiment \cite{Leebookchapter2020, Ackerman1966solidhelium, McNelly1970NaF}, a heat pulse is generated at the one end of the sample and the temporal response of temperature is monitored at the opposite end whereas light scattering techniques measure the local change of dielectric constants due to the propagation of second sound. At low temperatures, where phonon scattering is supposed to be dominated by the sample boundaries, the detector of the heat pulse experiments receives two temperature pulses corresponding the ballistic transport of transverse and longitudinal phonons. At little higher temperature, N scattering seems to dominate the phonon-phonon scattering and a distinct peak can be observed, which is the representative of second sound \cite{Jackson1970NaF, Jackson1971NaF, McNelly1970NaF}. At further elevated temperature, U scattering dominates and the second sound pulse broadens and gradually smears out into the diffusive signal. In their two consecutive studies, using the Ruggeri's model \cite{Ruggeri1990}, Tarkenton \textit{et al.} \cite{Tarkenton1994, Tarkenton1995} investigated the nonlinear wave propagation in solids and found the nonlinear corrections to the speed of the second sound for NaF and Bi are small. 

At earlier times, heat-pulse methods also suffered some disadvantages to detect second sound signal. One of the crucial restrictions involved the requirement of the absorption lengths of the order of the sample dimensions \cite{Pohl1976NaF}. To resolve these issues, in an alternative method, light scattering techniques probe second sound by measuring the local change of dielectric constants due to the propagation of second sound. The problem of weak coupling between light and thermal fluctuation at low temperatures had been resolved using force thermal scattering (FTS) technique \cite{Pohl1976NaF}. Both of these techniques detected second sound in NaF quite satisfactorily with reasonable agreement on the second sound speed and the temperature of occurrence \cite{Jackson1970NaF, Pohl1976NaF}. 

The fundamental difficulties in detecting second sound experimentally lies in the phenomena of coupling between temperature and other elementary excitations. However, it was found that SrTiO$_{3}$ possesses strong Normal scattering due to strongly anharmonic soft transverse optical phonons \cite{Gurevich1988}. Motivated by this idea, Koreeda \textit{et al.} \cite{Koreeda2007SrTio3, Koreeda2009SrTio3} explored low-frequency light scattering experiments without employing a thermal fluctuation field to investigate the propagation of second sound in SrTiO$_{3}$. They observed an underdamped second sound for SrTiO$_{3}$ below 40 K \cite{Koreeda2009SrTio3}. The origin of the anomalously broad Brillouin component was understood as the effect of second sound in SrTiO$_3$ in this Quasielastic Light Scattering (QELS) study. In recent times, advancements of the experimental techniques lead to more precise account of the second sound in solids. Khodusov \textit{et al.} \cite{Khodusov2009} observed a weakly damped second sound in the isotopically highly pure quantum crystals of orthodeuterium and parahydrogen as well as in the neon cryocrystals. Recently, Huberman \textit{et al.} \cite{Huberman2019graphite} experimentally observed second sound in graphite at moderately high temperature ($>$ 100 K) using transient thermal grating (TTG) technique implemented with time-resolved optical measurements. Very recently, Ding $\textit{et al.}$ \cite{Ding2022} observed second sound in graphite at even higher temperature ($>$ 200 K) using sub-picosecond TTG technique supported by first-principles simulations. For isotopically pure graphite, the occurrence of second sound had been predicted \cite{Ding2022} to reach even at room temperature. Due to the strict frequency bounds and the experimental limitations, second sound in solids had been experimentally explored mostly in a narrow temperature range. Recently, Beardo \textit{et al.} \cite{Beardo2021Ge} carried out an experiment with a rapidly varying temperature field as a driving force in a system of bulk Ge using a harmonic high-frequency external thermal excitation. High-frequency second sound was found for Ge in a wide temperature range (7 K - 300 K) observing the phase lag of the thermal response of the material and validated by $\textit{ab initio}$ and nonequilibrium MD approaches. These new experiments open up possibilities to explore the occurrence of second sound in a wide range of materials.

\subsubsection{Numerical investigations:}

Advancement of computational resources in late 80's and 90's opened up the avenues to explore the peculiar behavior of second sound in solids. It gathered more momentum in the post-2000 era due to the presence of large scale simulation tools for extremely time consuming atomistic and quantum mechanical methods like molecular dynamics and density functional theory etc. Also, computational resources greatly helped developing the extensive numerical solutions of the Peierls Boltzmann transport equation with lesser approximations on the phonon-phonon scattering processes. Several molecular dynamics studies \cite{Tsai1973, Tsai1976MD, Schneider1978MD, Osman2005MD, Shiomi2006MD, Kim2007MD, Yao2014MD, Zhang2011} were carried out in this context. Amongst them, The works of Tsai and MacDonald \cite{Tsai1973, Tsai1976MD} demands special attention as the molecular dynamics approach they adopted in the 70's were fundamentally very different from the then existing methods but it was surprisingly consistent with the theoretical results \cite{JosephHeatwaves1989}. Instead of linearizing the equations, they included the complete anharmonicity of the interatomic potential for forces. Their MD study \cite{Tsai1976MD} of an intense heat pulse propagation in a lattice at high temperature and pressure revealed the second sound propagation, superimposed on a diffusive background. A coupling between elastic and the thermal response was observed where longitudinal and transverse stress waves carry temperature waves with velocity resembling second sound velocity \cite{Tsai1976MD}. Another MD study \cite{Tsai1973} by the same authors on the propagation of shock wave in a 3D crystalline lattice revealed the existence of second sound in a thermally equilibrated regime behind the shock front. Schneider and Stoll \cite{Schneider1978MD} identified the temperature window and damping of second sound in model solid via the resonance in spectral density functions using a canonical MD ensemble with almost constant energy. Osman and Srivastava \cite{Osman2005MD} used MD simulations of heat pulse propagation in single-walled carbon nanotubes (SWCNT) and observed that the energy carried by wave packets corresponding to the second sound was larger compared to that of the twisted phonon mode (TW) and longitudinal acoustic (LA) modes. In another MD simulation of multiwalled CNT \cite{Kim2007MD}, however, the second sound feature was not seen. In a comparatively recent non equilibrium MD simulation of heat pulse propagation, Yao $\textit{et al.}$ \cite{Yao2014MD} observed an attenuated second sound propagating in both armchair and zigzag graphene. However, these MD simulations \cite{Osman2005MD, Shiomi2006MD} suffered from the space and time scales limitations. Later, using an optimized tersoff potential to account the atomic interactions in a lattice dynamics calculation of a (20, 20) SWCNT \cite{Lee2017}, those limitations were overcome and a significant contribution ($\geq$ $70$ $\%$) of the drifting phonons was identified in the heat propagation at room temperature. Also some of the non-equilibrium MD studies \cite{Zhang2011, Yao2014MD} underwent difficulties in detecting second sound due to its strict window condition and due to the small size and high temperature. A dispersion relation was also derived \cite{Lee2017, Leebookchapter2020} and the propagation and damping of second sound were understood in terms of the real and imaginary parts of the dispersion relation respectively.

Another approach within numerical methods dealt with modeling hydrodynamic phonon transport using approximate solutions \cite{YuReview2021} to the Boltzmann transport equations. In this context, the paradox of infinite propagation of speed of thermal signals in Fourier's law was overcome by using Cattaneo-Vernotte \cite{Cattaneo1948, Vernotte1958} or Guyer-Krumhansl (GK) \cite{GK1:1966} equation for heat conduction in solids. There are also other macroscopic equations for hydrodynamic heat conduction which is out of the scope of this review and readers are recommended to read the works of Guo and Wang \cite{Physrep2015Guo, Guo2018}. Very recently, Scuracchio $\textit{et al.}$ \cite{Scuracchio2019numerical} derived a system of coupled integrodifferential equations for phonon density fluctuations in 2D crystals and the second sound doublet was observed via the dynamic displacement susceptibility for 2D crystals. Using a discrete gas kinetic scheme, Luo $\textit{et al.}$ \cite{Luo2019} studied the propagation of second sound in graphene ribbon and found the flexural acoustic modes (ZA) of the phonon spectrum of graphene as the principal contributor to the second sound. Very recently, Shang $\textit{et al.}$ \cite{ShangSrep2020} obtained a 2D GK equations to describe hydrodynamic phonon transport and reached similar conclusions about the connections between ZA phonon modes and second sound. They found \cite{ShangSrep2020} the speed of second sound is much smaller than that of the Debye model at similar temperature due to the frequency dependent group velocity and the frequency independent phonon density of states of the ZA modes.

The evolution and expansion of first-principles techniques like Density functional methods \cite{Lindsay2019} with a lot of publicly available extremely efficient software packages (Quantum Espresso \cite{qe}, ShengBTE \cite{ShengBTE_2014}, PHONOPY \cite{phonopy}, PHONO3PY \cite{phono3py, chaput}) greatly helped the community to undertake the exploration into the  `No Man's land' in the field in terms of the computational feasibility. The first-principles calculations by Lee $\textit{et al.}$ \cite{Broido} predicted the existence of phonon hydrodynamics and therefore propagation of second sound in suspended graphene at higher temperatures and in wider temperature window compared to the 3D bulk materials. Cepellotti $\textit{et al.}$ \cite{Cepellotti2015} extended and generalized this phenomena at room temperature for a wide range of 2D materials using first-principles density-functional perturbation theory with an exact variational solution. Markov $\textit{et al.}$ \cite{Bitheory2018} carried out an exact variational solution to the BTE incorporating with the first-principles calculations of the three-phonon scattering and consistently identified the drift velocity of Bi along the binary axis with that of the second sound in Bi measured experimentally \cite{Narayanmurti1972Bi} many years back. Very recently, hydrodynamic features in bulk crystalline polymers were also observed by Zhang and co-workers \cite{ZhangPolymer2020}. Also, very recently, Monte-Carlo simulations \cite{LeeMC2019, NieMC2020} and Lattice-Boltzmann method \cite{GuoWangLB2022} were introduced to predict second sound in solids. Apart from the exact solution of BTE \cite{chaput}, Macroscopic hydrodynamic equations were also coupled with the first-principles calculation inputs of the harmonic and anharmonic properties of the crystal lattice to predict the phonon hydrodynamics in materials. Kinetic collective model (KCM) \cite{Torres2017KCM, Thomas2014KCM, Thomas2015KCM, Alvarez2009KCM, alvarez2018thermalbook, Beardo2019KCM, Melis2019KCM} is one such model derived from the Guyer and Krumhansl solution \cite{GK1:1966} to the LBTE which splits the collision operator into Normal and resistive ones and thus separates kinetic and collective contributions of phonons to the heat conduction. This model also uses first-principles outputs of second and third order force constants due to harmonic and anharmonic processes of phonons. This can be a viable alternative for some instances where the complete solution of LBTE with first-principles force constants calculations demand unprecedented computational resources. KCM in conjunction with Guyer-Krumhansl frequency criteria, was employed in some recent studies to predict second sound and phonon hydrodynamics in some 3D materials \cite{kanka2, Torres_2019, kanka3}.

\subsubsection{Second sound from Relaxon approach:}

In the realm of relaxon approach, introduced by Cepellotti and Marzari \cite{Transport_waves_as_crystal_excitations, Cepellotti2016}, linear superpositions of phonon modes are termed as Relaxons which represent the collective excitation in the solid having well defined lifetimes and mean free paths. It was shown that the LBTE naturally allows the existence of this collective excitations contrary to the earlier works \cite{GK1:1966, Hardy1970} where the solutions of LBTE were associated with approximations and simplifications. Also the single relaxation time approximation and Debye approximations for phonon dispersion were abandoned. The relaxon approach based on the fact that the temperature waves are related to the fluctuations of the phonon population of the solid. Under scattering events, this population deviates from the Bose-Einstein to a displaced Bose-Einstein distribution, inducing a change of the total energy of the crystal. This total energy fluctuations is related to the temperature fluctuation in the system through the specific heat of the material. Cepellotti and Marzari employed LBTE for the \cite{Transport_waves_as_crystal_excitations} 

\begin{widetext}
\vspace*{-2.9cm}
\begin{figure}[H]
    \centering
\includegraphics[width=0.8\textwidth]{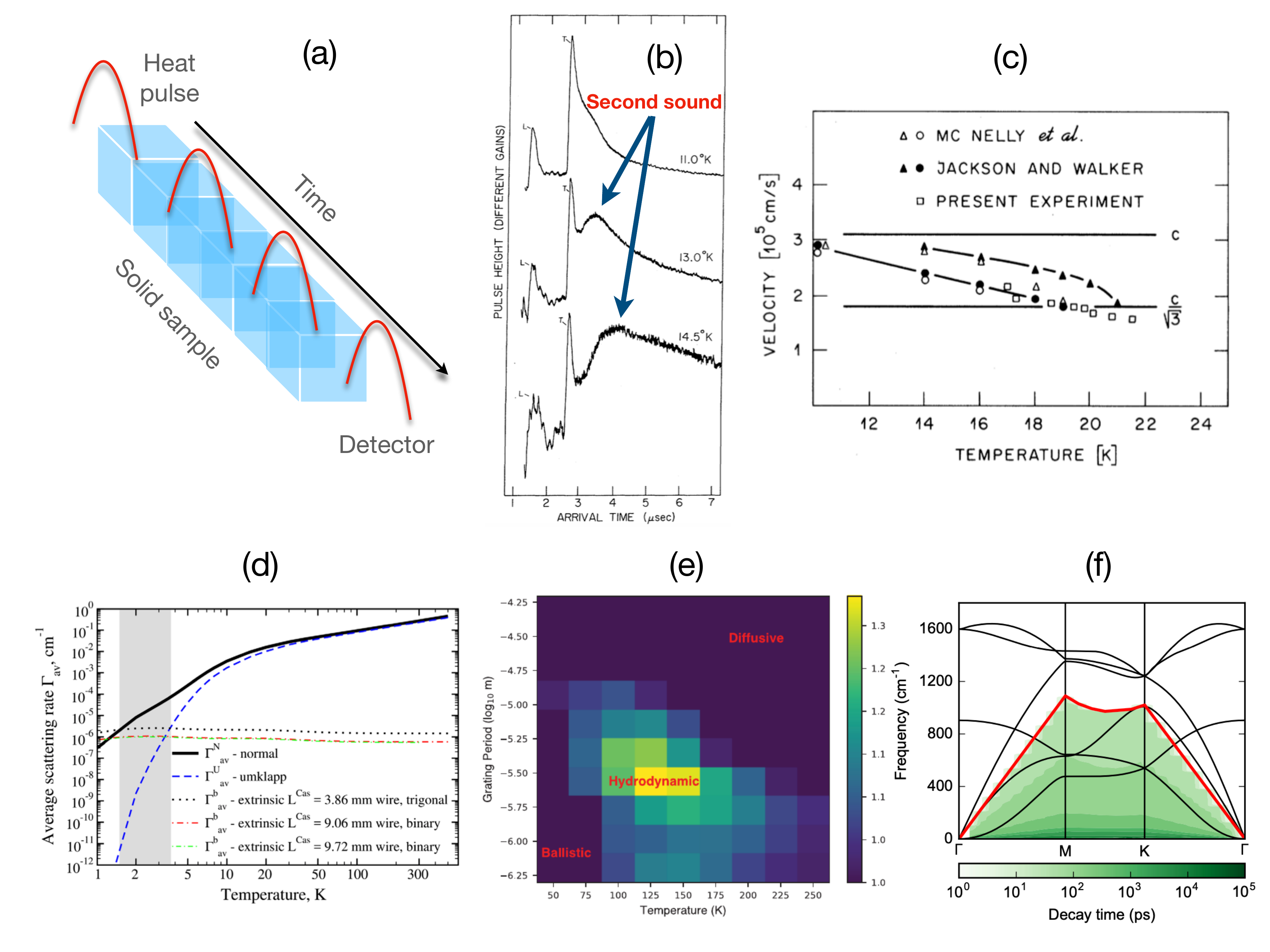}
\vspace*{-1.0cm}
\caption{\footnotesize Different observations of Second sound in crystalline solids. (a) Schematic of second sound propagation in a solid. A heat pulse, introduced at one end, can be received at the detector at the other end with negligible damping. (b) Results from the observation of second sound in heat pulse experiments of NaF, presented via pulse heights as a function of arrival times of the pulses. The pulse emerging from the shoulder of the curves after the longitudinal (L) and transverse (T) ballistic peaks, denotes the second sound signal. Figure \ref{fig:second_sound}. (b) adapted with permission from Ref. \cite{Jackson1971NaF}. Copyrighted by the American Physical Society. (c) Comparison between different experiments to observe first and second sound velocities in NaF as a function of temperature. The results of heat pulse \cite{Jackson1971NaF} and the evidences from light scattering experiments \cite{Pohl1976NaF} are shown to be consistent for NaF. Reprinted (figure) with permission from Ref. \cite{Pohl1976NaF}. Copyright (1976) by the American Physical Society. (d) Temperature variation of the average scattering rates corresponding to N, U and B scattering for Bi calculated via $\textit{ab initio}$ techniques. The shaded regime denotes the GK criterion for the occurrence of second sound. Reprinted (figure) with permission from Ref. \cite{Bitheory2018}. Copyright (2018) by the American Physical Society. (e) Second sound window of graphite with natural isotope content is shown in the temperature and the TTG grating period parameter space with the color code denoting the ratio of the maximum at the peak of the magnitude of frequency-domain Green's functions to the minimum between the peak and zero frequency as defined in \cite{Huberman2019graphite}. Reprinted with permission from Ref. \cite{Huberman2019graphite}, AAAS. Republished with permission of [``CCC"], from \cite{Huberman2019graphite}; permission conveyed through Copyright Clearance Center,Inc. (f) Continuous spectrum of dispersion relations for temperature waves in graphene in the `relaxon' picture, over a high symmetry path in the Brillouin zone, is presented. The largest value of the allowed oscillation frequencies are marked with red while zero temperature phonon dispersion curves are denoted via black. The color scale implies decay time of the temperature waves. Reprinted (figure) with permission from Ref. \cite{Transport_waves_as_crystal_excitations}. Copyright (2017) by the American Physical Society.}

    \label{fig:second_sound}
\end{figure}

\end{widetext}

\noindent displaced phonon distribution and arrived at an eigenvalue equation where eigenvectors of the scattering matrix correspond to the crystal excitations. Investigating graphene, the authors reached the conclusion that transport waves exist in many crystals but the observation demands to meet the criteria of long relaxation times and frequency-resolved advanced experimental methods. Another work by Simoncelli $\textit{et al.}$ \cite{Simoncelli2020} exploited the relaxon's even parity to describe a generalized viscous equations and the second sound was explored using this viscous heat equations.

Figure \ref{fig:second_sound} summarizes the exploration of second sound in different materials using various experimental, theoretical and numerical methods over the years and illustrates some of the crucial findings in the literature concerning second sound in solids.

\subsection{\textbf{Poiseuille flow}}

Phonon Poiseuille flow is another exotic phonon hydrodynamic phenomena in solids which bears resemblance with the Poiseuille flow of fluids in a pipe. This phenomenon operates in a thermal conduction regime where Normal scatterings are predominant and the thermal resistance is introduced by the boundaries of the sample. The abundance of N scattering events causes the deviation from the Bose-Einstein distribution of phonons with a drift velocity developing along the direction of the thermal gradient. However, diffuse boundary scattering events tend to lower the drift velocity of phonons at the boundaries, giving rise to a drift velocity gradient normal to the heat flow direction. The steady-state phonon hydrodynamical feature where phonons flow under a thermal gradient with a drift is termed as Poiseuille flow \cite{Broido} as it bears similarities with the fluid flow in a pipe where the pressure gradient plays similar role as that of the temperature gradient and the flow resistance comes from the viscosity and the pipe diameter, comparable to the N scattering rates and sample width respectively in the context of thermal conduction \cite{CahillPohl1988}. The drift velocity gradient across the width induces a phonon momentum transfer towards the boundary from the center of the width. N scattering events hinder this cross-plane momentum transfer and thus invokes the thermal resistance as viscous damping effect \cite{Leebookchapter2020}. Thus, similar to the concept of fluids, N scattering along with the boundary scattering give rise to the idea of hydrodynamic viscosity of phonons, which has also been realized via viscous heat equations in relaxon approach \cite{Simoncelli2020}. Similar to the second sound propagation regime, Poiseuille flow does not exist in either ballistic or diffusive (or kinetic) regimes as the sample size is less and much greater than the average phonon mean free path in these two regimes, respectively. In the first scenario, the heat conduction is ballistic as the phonons directly hit the boundaries before encountering other phonons. In the second scenario, much larger system size compared to the phonon mean free paths allows resistive phonon-phonon scattering events (Umklapp) to become the principal mechanism for thermal resistance inside the material, defining the diffusive thermal conduction regime. These consequences bring forth the idea of a mean free path window in which the Poiseuille flow can be realized.

\subsubsection{Theoretical predictions:}

The investigation on the thermal conductivity of a crystal at low temperature by Sussmann and Thellung \cite{Sussmann1963} was the first to identify the Poiseuille like flow in a phonon gas, neglecting the Umklapp scattering at low temperatures. A temperature gradient was set up between the two ends of a cylinder with a rough boundary. The hydrodynamic equations, derived for the phonon gas, coupled with mean free approximation and ignoring dispersion, was shown to feature two distinct contributions to the heat flow: (a) one due to the drift motion of the phonon gas and (b) one due to the temperature gradient. This drift was eventually shown to produce Poiseuille flow. Also, using a simple substitution of the drift velocity, the coupled equations derived for temperature and drift velocity was seen emerging as undamped temperature wave equation corresponding to the second sound if the thermal dissipation was neglected \cite{Sussmann1963}. Following Sussmann and Thellung \cite{Sussmann1963}, a very brief derivation of Poiseuille flow is given below. The coupled equations for both temperature and drift velocity, as was obtained by Sussmann and Thellung \cite{Sussmann1963} reads
\begin{equation}
    \frac{\dot{T}}{T_0} + \frac{1}{3}\nabla \cdot \mathbf{u} - \frac{1}{3}C_{l}^{2}C_{t}^{2} \frac{\tau_{l} C_{t}+ 2\tau_{t}C_{l}}{C_{t}^{3}+2C_{l}^{3}}\frac{\Delta T}{T_0} = 0
\end{equation}
and 
\begin{equation}
    \dot{\mathbf{u}} + C_{l}^{2}C_{t}^{2}\frac{C_{t}^{3}+2C_{l}^{3}}{C_{t}^{5}+2C_{l}^{5}}\frac{\nabla T}{T_0}-C_{l}^{2}C_{t}^{2}\frac{\tau_{l}C_{t}^{3}+2\tau_{T}C_{l}^{3}}{C_{t}^{5}+2C_{l}^{5}}\left[ \frac{2}{5}\nabla (\nabla \cdot \mathbf{u})+\frac{\nabla u}{5}\right] =0
\end{equation}

\noindent Considering heat flow in a cylinder whose length is much larger than the radius and assuming diffusive boundary scattering at the cylindrical surface, retaining only axial components of $\nabla T$ and the drift velocity \textbf{u} (along $z$ axis) reduce these above two equations as 

\begin{equation}{\label{drift}}
    \left(\frac{1}{C_{l}^{3}}+\frac{2}{C_{t}^{3}}\right)u_{z}-\frac{1}{T_0}\nabla_{z} T\left(\frac{\tau_l}{C_l}+\frac{2\tau_t}{C_t}\right) = f\left(x,y\right)
\end{equation}

and

\begin{equation}{\label{temp}}
    \frac{1}{T_0}\nabla_{z} T - \frac{\tilde{\tau}}{5}\left(2\nabla_{z}^{2}u_z+\Delta u_z\right) = 0
\end{equation}
with $\tilde{\tau}$ = $\frac{\tau_{l}C_{t}^{3}+2\tau_{t}C_{l}^{3}}{C_{t}^{3}+2c_{l}^{3}}$. As Sussmann and Thellung \cite{Sussmann1963} approached a dispersion-less low temperature phonons, their longitudinal and transverse energies were realized using the proportionality with momentum, giving rise to the constants $C_l$ and $C_t$ with $\tau_l$ and $\tau_t$ their mean free times respectively. The solutions of these solutions constitute Poiseuille flow of phonons with a parabolic nature of the drift velocity, 

\begin{equation}
    \nabla_{z}T = constant
\end{equation}
and 
\begin{equation}
    u_{z}\left(r\right) = \frac{5\nabla_{z}T}{4\tilde{\tau} T_0} \left(r^{2}-R^{2}\right)
\end{equation}
where $R$ is the radius of the cylinder and $r$ = $\sqrt{x^{2}+y^{2}}$. The signature of the Poiseuille flow through the dependence of thermal conductivity on temperature and characteristic size was first realized by Gurzhi \cite{Gurzhi1964, Gurzhi1968}. A hydrodynamic equation was employed to solve LBTE by  series expansion in the small parameters $l_{N}/d$ and $l_{N}/l_{R}$ where $l$ is the mean free path, $d$ is the diameter of the sample and $N$ and $U$ denote Normal and Umklapp scattering respectively. Thermal conductivity in the rarefied Poiseuille flow regime ($l^N$ $\ll$ $d$ $\ll$ $l^U$), was found to exhibit a much stronger temperature dependence $\left(\kappa \propto d^{2} T^{8}\right)$ compared to that of the Casimir effect \cite{Casimir1938}. In 1966, a series of papers \cite{Krumhansl1965, GK1:1966, GK2:1966} extensively discussed and derived LBTE for the nonmetallic crystals from the analysis of the eigenvectors of the normal scattering operator using the relaxation time approximation. The existence of the Poiseuille flow was found to be correlated with the propagation of second sound \cite{GK2:1966, BeckReview1974} as both of them require the presence of drifting distribution of phonons. Therefore, the condition 
\begin{equation}
    \Gamma^{U} \ll \Gamma^{B} \ll \Gamma^{N}
\end{equation}
\noindent denotes the frequency window where both Poiseuille flow and second sound can operate. Guyer and Krumhansl \cite{GK2:1966} also identified another hydrodynamic regime where the Normal scattering processes are still dominant yet the heat flux is dissipated via Umklapp resistive scattering contrary to the Poiseuille regime where the sample boundary dominantly dissipates the heat. This regime was termed as Ziman hydrodynamic regime where the frequency window satisfies the following 
\begin{equation}
    \Gamma^{B} \ll \Gamma^{U} \ll \Gamma^{N}
\end{equation}
where $\Gamma$ denotes the average scattering rate with $i$ = $N$, $U$ or $B$ defining normal, Umklapp and boundary scattering processes respectively.
\begin{equation}
    \Gamma^{i} = \left\langle\tau_{i}^{-1} \right\rangle_{ave} =  \frac{\sum_{\lambda}C_{\lambda}\tau_{i\lambda}^{-1}}{\sum_{\lambda}C_{\lambda}}
\end{equation}
\noindent Here $C$ denotes the specific heat and $\lambda$ defines the phonon mode comprised of wave number and phonon branch. Further, invoking Poiseuille flow condition, Guyer and Krumhansl derived the temperature dependent expression for thermal conductivity which is consistent with that of the findings of both Gurzhi \cite{Gurzhi1964} and Sussmann $\textit{et al.}$ \cite{Sussmann1963}. Meier \cite{Meier1969} reproduced the results on Poiseuille flow derived by Guyer and Krumhansl using Green's functions approach. The thermal conductivity was found \cite{Meier1969} to vary with $T^8$, consistent with that of the Gurzhi. Nielsen \cite{Nielsen1969} found a nonlinear dependence of the heat current on temperature gradient under the Poiseuille flow conditions of a phonon gas. In another theoretical work \cite{Bausch1972}, Callaway's relaxation time approximation \cite{Callaway1959} was used to reconstruct LBTE to obtain numerical solution of the thermal conductivity for thin plates by an approximation of the integral equation for drift velocity. It was shown for model calculations of LiF and NaF that the Poiseuille flow could be visible only in a very pure crystals of small dislocation density and with more thickness \cite{Bausch1972}. Beck \cite{Beck1975} derived the temperature dependence of thermal conductivity by taking a model wave vector-dependence of Normal phonon scattering rate and observed that $\kappa$ $\propto$ $T^6$. Employing Cattaneo model \cite{Cattaneo1948} with adding nonlocal effects and with comparing with GK equation \cite{GK1:1966}, Jou $\textit{et al.}$ \cite{JOU199047} theoretically derived the consequences of generalized temperature on the effective thermal conductivity in the Poiseuille flow regime.

\subsubsection{Experimental realizations:}

After the theoretical prediction by Gurzhi \cite{Gurzhi1964, Gurzhi1968}, experimental efforts were made to observe the phonon Poiseuille flow in crystalline solids. All these experiments \cite{Mezhov-Deglin1965, Thomlinson1969, Kopylov1971, Kopylov1973, Seward1969, Hogan1969, Armstrong1979, Smontara1996, Zholonko2006, Inyushkin2004, Machida2018P, MartelliStrontium2018, Machida309} were based on measuring the thermal conductivity ($\kappa$) as a function of temperature and observing the temperature scaling of the $\kappa (T)$ below the $\kappa (T)$ peak. In 1965, for the first time, this hydrodynamic feature was experimentally shown by Mezhov-Deglin \cite{Mezhov-Deglin1965} for crystalline He$^4$. The thermal conductivity was found to vary with temperature as $\kappa (T)$ $\sim$ $T^n$, with $n$ varying from 6-8. Also, the maximum average mean free path was seen to exceed the diameter of the sample. These two hallmarks of Poiseuille flow was found to be consistent with that of the Gurzhi's theoretical work \cite{Gurzhi1964}. Thomlinson \cite{Thomlinson1969} evidenced a similar trend of $\kappa (T)$ with temperature with the exponent ($n$) ranging from 3.4 to 3.7. Poiseuille flow was also investigated at low temperature for hexagonal closed-packed He$^4$ \cite{Armstrong1979, Seward1969, Hogan1969}, quasi-one dimensional single crystals \cite{Smontara1996}, single crystals of Si \cite{Inyushkin2004} and for solid parahydrogen \cite{Zholonko2006}. Isotopically and chemically pure materials were observed to be more prone to display Poiseuille flow conditions \cite{Inyushkin2004}. Kopylov $\textit{et al.}$ \cite{Kopylov1971, Kopylov1973} explored this feature in pure Bi single crystals and in the temperature range between $T$ = 1.3 K to 2.5 K, $\kappa (T)$ was found to scale with $T^{3.15\pm 0.07}$. This exponent along with the corresponding growth in the effective mean free path ($l_{eff}$) with temperature, demonstrated the existence of Poiseuille flow in Bi. Recently, Machida $\textit{et al.}$ \cite{Machida2018P} observed a faster than cubic dependence of $\kappa (T)$ on $T$ for black Phosphorus in a temperature range of 5-12 K. The momentum exchange between acoustic phonon branches of black P was found to be responsible to felicitate the Poiseuille flow. Here we note the discrepancies between the exponents of the temperature of theoretical and experimental observations in the Poiseuille flow regime. Almost in all the experiments, the exponents were found to be less than that of the Gurzhi's \cite{Gurzhi1964, Gurzhi1968} findings. The reason behind these discrepancies comes from the understanding of the kinematic viscosity of the phonon system, emerging out of the N scattering and local velocity of phonons, as was mentioned by Gurzhi \cite{Gurzhi1968}. The variation of this phonon viscosity at a given temperature makes the phonon system non-Newtonian, giving rise to a comparatively flatter parabolic velocity profile which eventually lead to the absence of the superlinear size dependence of the thermal conductivity \cite{Machida2018P}. Martelli $\textit{et al.}$ \cite{MartelliStrontium2018} carried out experiments on both undoped and doped SrTiO$_3$ and Poiseuille flow was realized in the undoped sample via the temperature dependence of $\kappa (T)$ with an exponent $n$ $>$ 3 in the low temperature (6 K $<$ $T$ $<$ 13 K). Very recently, thin graphite sample was also shown to feature Poiseuille flow at reasonable high temperature (40 K) \cite{Machida309}. 

\subsubsection{Numerical explorations:}

The solution of BTE in the Guyer-Krumhansl approach \cite{GK1:1966} greatly helped the community to carry out numerical studies on the phonon hydrodynamics. Moreover, using dispersion relation of nonmetallic solids, Guyer and Krumhansl \cite{GKdispersion1964, GK2:1966} established the link between the occurrence of both Poiseuille flow and second sound in the same frequency window. Thus, GK approach was numerically adopted in many studies \cite{Jou2010, Sellitto2015, ShangSrep2020}, to realize the Poiseuille flow in phonon dynamics of the solids. Sellitto $\textit{et al.}$ \cite{Sellitto2015} analysed the nature of the heat flux profiles across a narrow 2D strip using GK-type generalized heat transport equation with a slight modification of the boundary conditions in the wall. It was found that only a small range of temperature and strip width is permissible for the Poiseuille flow of phonons with a parabolic heat flux profile \cite{Sellitto2015}. Also, superlinear dependence of heat current on the ribbon width was numerically understood as a manifestation of the Poiseuille flow in 2D materials \cite{ShangSrep2020}. Apart from that, different methods were performed to numerically observe the existence of the Poiseuille flow. These approaches include macroscopic heat conduction models \cite{DONG2014256, Guo2018}, hydrodynamic models concomitant with 2D crystals \cite{Scuracchio2019numerical, XU2019126017}, first-principles density functional calculations coupled with full solution of LBTE using either variational \cite{Cepellotti2015, Bitheory2018} or iterative methods \cite{Broido, Ding2018}, employing second-principles polynomial potential \cite{Torres2019} with the direct solution \cite{chaput}, solution under Callaway model \cite{Guo2017} and Monte Carlo solutions of Peierls BTE \cite{Li2018, LeeMC2019, NieMC2020} etc. Similar to what has already been discussed in the second sound observation in earlier section, first-principles calculations with accurate solution of LBTE using different methods \cite{chaput, Cepellotti2015, Broido} helped a lot to predict the correct thermal conductivity and therefore the Poiseuille like behavior in 2D materials. At room temperature, phonon hydrodynamics seems crucial for graphene and a wide temperature \cite{Cepellotti2015} and width windows \cite{Broido} were observed for the occurrence of the Poiseuille flow of phonons. Further, Callaway's model \cite{Callaway1959} was broadly found to predict the correct behavior even for the 2D materials \cite{Cepellotti2015}. Li $\textit{et al.}$ \cite{Li2018} employed a deviational Monte Carlo scheme \cite{HadjiMC} coupled with first-principles scattering matrices due to anharmonicity to study the phonon hydrodynamics in suspended graphene. This novel numerical scheme, introduced by Landon and Hadjiconstantinou \cite{HadjiMC}, deals with the sampling of the deviation of the distribution function by attaching either positive or negative values of unit deviational energies attached with each of the particles. These particles, however, are not to be mixed with real atoms or molecules and it had been approximately described as the constitutive computational elements in the distribution function \cite{HadjiMC}. Li $\textit{et al.}$ \cite{Li2018} showed that this Monte Carlo technique is at par with the other methods to solve LBTE and found a superlinear dependence of the thermal conductivity on the width, confirming the existence of Poiseuille flow at 100 K and in the width-window between 1-10 $\mu$m of graphene. Here, we recall the differences seen between Gurzhi's theoretical model and experimental realizations as was discussed in the previous subsection. Employing $\textit{Ab initio}$ calculations coupled with the iterative solution of LBTE, phonon hydrodynamic investigations by Ding $\textit{et al.}$ \cite{Ding2018} identified the Poiseuille flow in graphite using the superlinear size dependence of the thermal conductivity, consistent with the prediction by Gurzhi \cite{Gurzhi1964, Gurzhi1968}. The thickness dependence of thermal conductivity was found to be either superlinear or sublinear depending on the variation of the boundary scattering rate compared to the normal scattering rates. Recently, in the hydrodynamic phonon transport in GeTe at low temperature, an $\textit{ab initio}$ numerical study \cite{kanka3} also suggested a similar superlinear size dependence and demonstrated it using an exponent related to the ratio between normal and resistive scattering rates.

\subsubsection{Relaxon approach to Poiseuille flow:}

As mentioned earlier, the relaxon picture \cite{Cepellotti2016} approaches the phonon hydrodynamics from the inconsistent description of the Fourier's law and is based on the collective phonon excitations (superposition of phonons) instead of individual phonon dynamics. Using the idea of relaxon, Simoncelli $\textit{et al.}$ \cite{Simoncelli2020} derived generalized viscous heat equations involving two coupled equations for local temperature and the drift velocity fields, which on the limiting conditions of crystal momentum dissipation, invoke either second sound (weak dissipation) or Fourier's law (strong dissipation). Rewriting the viscous heat equations as energy and momentum balance equations, heat flux was realized \cite{Simoncelli2020} as separate effects coming from the temperature driven ($\textbf{Q}^{\delta} (\textbf{r},t)$) and the drift velocity driven ($\textbf{Q}^{D} (\textbf{r},t)$) components. Employing no-slip boundary conditions (zero drift velocity of the relaxons), the heat flux profiles were found to feature Poiseuille flow associated with a characteristic length scale which dictates the parabolic variation of the heat flux \cite{Simoncelli2020}. These findings were found to be consistent with that of the space-dependent LBTE solution using full scattering matrix \cite{Andrea2017}. The analytical solutions of the one-dimensional version of the viscous heat equations  

\begin{figure}[H]
    \centering
\includegraphics[width=0.76\textwidth]{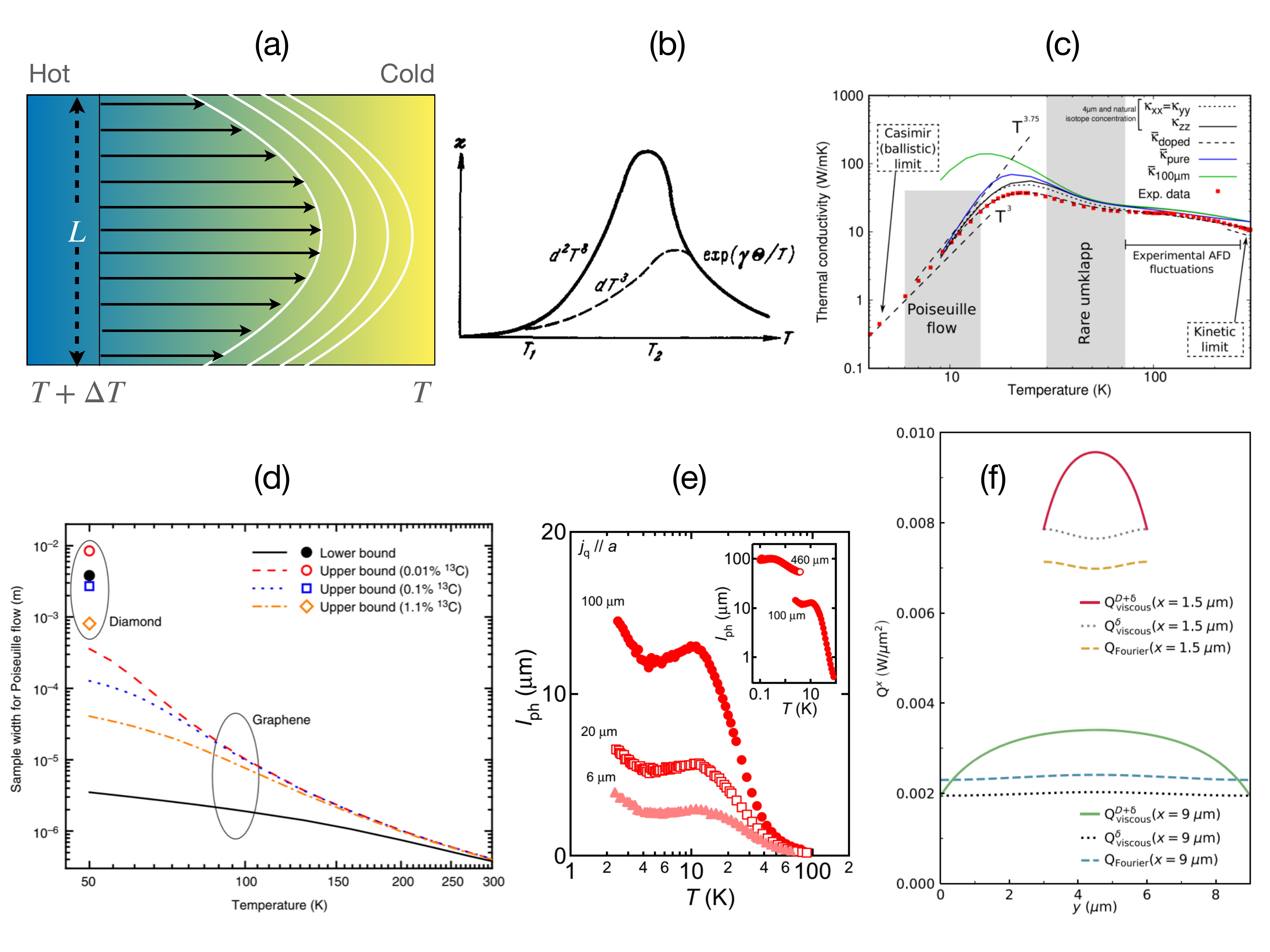}
\vspace*{-1cm}
    \caption{\footnotesize Different instances of Poiseuille flow of phonons in crystalline solids. (a) A schematic diagram of phonon Poiseuille flow inside a solid sample of finite width. The drift motion is larger at the center compared to the boundaries, giving rise to a parabolic flow pattern of the heat flux profile. (b) Thermal conductivity as a function of temperature is shown to vary with stronger temperature dependence ($\propto$ $T^8$) than the Casimir (ballistic) limit ($\propto$ $T^3$) in the Poiseuille flow regime. Reprinted with permission from Ref. \cite{Gurzhi1968}, Russian Academy of Sciences. $\copyright$ Uspekhi Fizicheskikh Nauk 1968. (c) The temperature variation of the thermal conductivity of crystalline SrTiO$_3$ is presented employing LBTE coupled with $\textit{ab initio}$ density functional simulations. A faster than $T^3$ dependence ($\kappa$ $\propto$ $T^{3.75}$) is observed in the Poiseuille flow regime. Reprinted (figure) with permission from Ref. \cite{Torres2019}. Copyright (2019) by the American Physical Society. (d) First-principles findings of the sample width window as a function of temperature to detect Poiseuille flow in graphene and diamond. Graphene has a wider and distinct gap below 100 K compared to the negligible gap in diamond (even at 50 K), making it more prone to the Poiseuille flow characteristics. Reprinted by permission from Springer Nature Customer Service Centre GmbH: [Springer Nature] [Nature Communications] \cite{Broido} (2015). (e) Effective phonon mean free path ($l_{ph}$) is extracted from $\kappa$ as a function of $T$ along the $a$-axis of black phosphorus. The peak in the $T$ dependence of $l_{ph}$ indicates Poiseuille flow which decreases with sample width. Reprinted/adapted from \cite{Machida2018P}. $\copyright$ The Authors, some rights reserved; exclusive licensee AAAS. Distributed under a CC BY-NC 4.0 license \url{http://creativecommons.org/licenses/by-nc/4.0/} (f) Poiseuille flow in `relaxon' picture using viscous heat equations. $x$ component of the heat flux along the sections $x$= 1.5 and 9 $\mu m$ are presented for graphite. Unlike the Fourier heat flux, total heat flux obtained from viscous heat equations show Poiseuille-like profile. Reprinted with permission from Ref. \cite{Simoncelli2020}. \href{https://creativecommons.org/licenses/by/4.0/}{CC BY 4.0}.}
    \label{fig:Poiseuille_flow_summary}
\end{figure}

\noindent were also seen to produce similar qualitative behavior with the findings of Sussmann and Thellung \cite{Sussmann1963}. Some of the important results from the literature that feature phonon Poiseuille flow in crystalline materials using various experimental, theoretical and numerical approaches are presented in Figure \ref{fig:Poiseuille_flow_summary}.

\subsection{\textbf{Knudsen minimum}}

Citing the analogy of phonon flow with the Knudsen flow of gas, phonon Knudsen minimum is ascribed to the minimum in temperature dependence of the phonon mean free path, marking ballistic to hydrodynamic phonon transport. It indicates a minimum in the normalized heat flow rate in a system whose size becomes comparable to the phonon mean free path.

\noindent In comparison with the second sound and Poiseuille flow, phonon Knudsen minimum had been demonstrated and discussed less often in the literature in the context of phonon hydrodynamic heat conduction. This is primarily because the occurrence of the phonon Knudsen minimum is more subtle compared to its two other counterparts as it corresponds to the heat conduction regime where the transition between ballistic and hydrodynamic regimes is occurred \cite{GangChenbook, Lindsay2019}. Here we note an important point that despite the GK conditions \cite{GK2:1966} with a quantifiable frequency window exist to detect the phonon hydrodynamic regimes in a crystal, often the boundaries between different heat conduction regimes are blurred \cite{kanka3} (ballistic-hydrodynamic, hydrodynamic-diffusive, ballistic-diffusive). This emerges from the GK condition \cite{GK2:1966} for the existence of the Poiseuille hydrodynamics: $\Gamma^R$ $\ll$ $\Gamma^B$ $\ll$ $\Gamma^N$, where $\Gamma$ denotes average scattering rates and $R$, $B$, $N$ stand for resistive (Umklapp and isotope scattering), boundary and normal scattering processes respectively. This inequality prevents from defining a sharp boundary between different regimes. For example, starting from the ballistic regime, the boundary between the ballistic ($\Gamma^R$ $\ll$ $\Gamma^N$ $\ll$ $\Gamma^B$) and the hydrodynamic heat conduction ($\Gamma^R$ $\ll$ $\Gamma^B$ $\ll$ $\Gamma^N$) goes through a crossover where the condition shifts from $\Gamma^N$ $<$ $\Gamma^B$ to $\Gamma^N$ $>$ $\Gamma^B$ where the difference between $\Gamma^N$ and $\Gamma^B$ is not significant to satisfy the GK conditions.

The microscopic understanding of the phonon Knudsen minimum in solids relies on the occurrence of a minimum in the normalized heat flow rate in a system of which the characteristic size ($L$) becomes comparable to the phonon mean free path (MFP) \cite{Ding2018, Broido}, in other words when Knudsen number (Kn = MFP/$L$) is close to 1. Further, it has been understood by considering the minimum in the variation of the dimensionless thermal conductivity ($\kappa^{*}$) as a function of the sample width or the characteristic size. In the complete ballistic regime, where system size defines the mean free path of phonons, thermal conductivity and $\kappa^{*}$ are seen varying linearly and remaining constant with the system size, respectively. When N scattering is introduced, the thermal resistance is solely controlled by the boundary scattering and therefore it depends on the system size. For small characteristic size of the sample N scattering is seen to increase the boundary scattering and lower the $\kappa^{*}$ than the ballistic case. On the other hand, larger size weakens the boundary scattering leading to a larger $\kappa^{*}$ than the ballistic case. Therefore, $\kappa^{*}$ goes through a minimum in between these two situations, termed as Knudsen minimum. Larger $\kappa^{*}$ than the ballistic case, is an indicative of a superlinear size dependence of the thermal conductivity. Therefore, Knudsen minimum occurs concomitantly with the onset of the Poiseuille flow in solids and more specifically, a Poiseuille peak in phonon hydrodynamic regime is followed by a Knudsen minimum.

\begin{figure}[H]
    \centering
\includegraphics[width=1.0\textwidth]{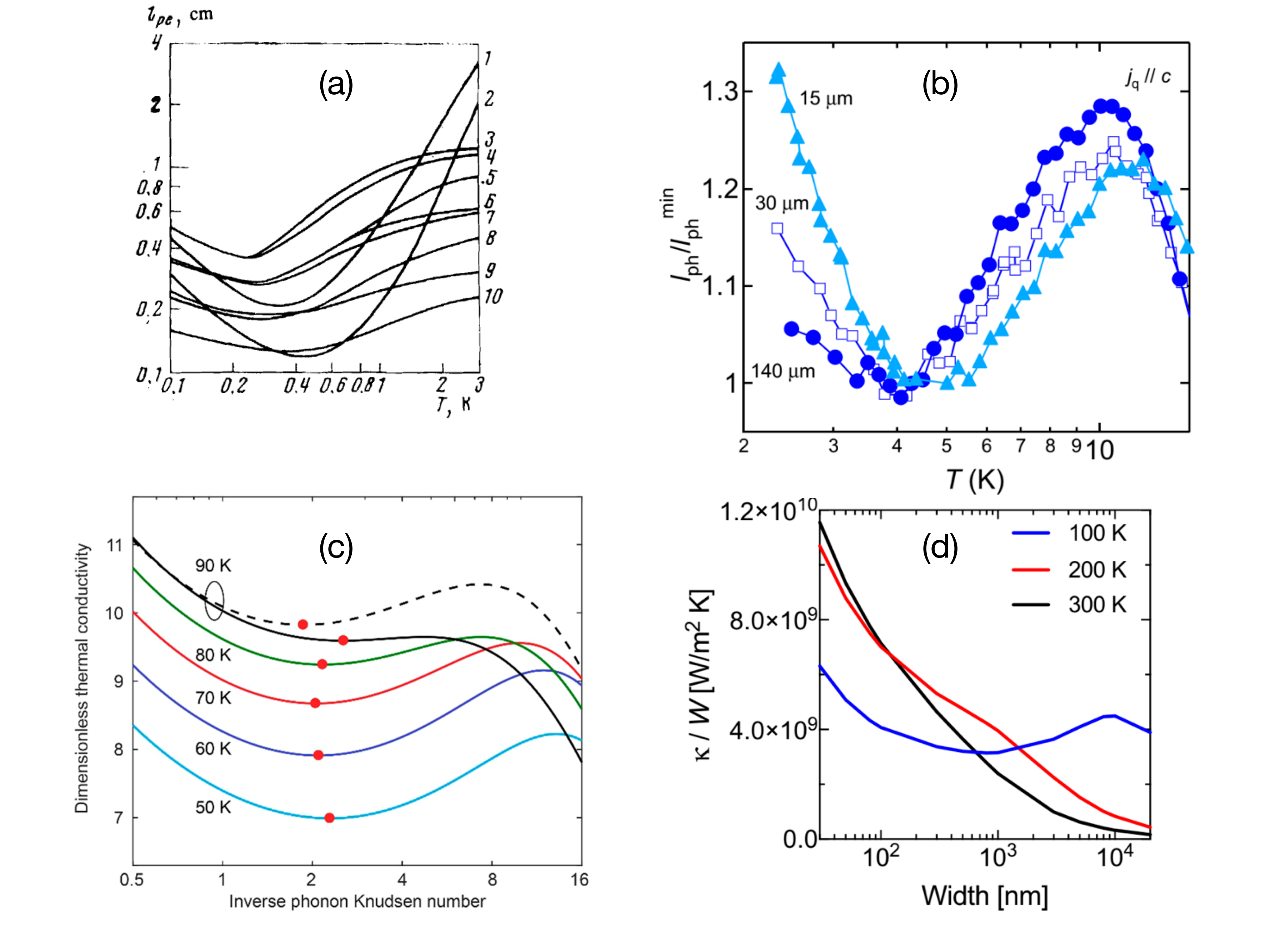}
\vspace*{-1cm}
    \caption{Knudsen minimum in crystalline solids. (a) Experimentally observed Knudsen minimum in Bi crystal below 2 K, shown via the variation of effective mean free path ($l_{pe}$) with temperature. Reproduced from \cite{Mezhov1975}, Russian Academy of Sciences. $\copyright$ Journal of Experimental and Theoretical Physics. (b) Thickness dependence Knudsen minimum of black phosphorus, expressed via the temperature variation of the effective phonon mean free path along c axis, normalized by its value at Knudsen minimum. With increasing thickness, minimum shifts to lower temperatures. Reprinted/adapted from \cite{Machida2018P}. $\copyright$ The Authors, some rights reserved; exclusive licensee AAAS. Distributed under a CC BY-NC 4.0 license \url{http://creativecommons.org/licenses/by-nc/4.0/} (c) Simulation of Poiseuille flow and Knudsen minimum in graphite ribbon, realized via minimum in nondimensional thermal conductivity (red dots) while varying with inverse phonon Knudsen number. Reprinted (figure) with permission from Ref. \cite{Ding2018}. Copyright (2018) American Chemical Society. (d) Knudsen minimum was found in the width variation of thermal conductivity of suspended graphene, normalized by sample width at different temperatures via solving the Peierls BTE with deviational Monte Carlo scheme. Knudsen minimum seems prominent when temperature is lowered from 300 K to 100 K. Reprinted (figure) with permission from Ref.  \cite{Li2019}. Copyright (2019) American Physical Society.}
    \label{fig:knudsen_minimum}
\end{figure}

After the exploration of molecular Knudsen minimum observed by Knudsen \cite{Knudsen1909} in early 20th century, Cercignani and Daneri \cite{Cercignani1963} was the first to numerically solve the Boltzmann transport equation for the Poiseuille flow of a rarefied gas between two parallel plates and found the Knudsen minimum in the variation of nondimensional volume flow rate with inverse Knudsen number. Liquid helium was shown to exhibit Knudsen minimum at very low temperature \cite{Whitworth1958}. The solids, on the other hand had not been found to be very prone to display the phonon Knudsen minimum \cite{GangChenbook}. In 1975, Mezhov-Deglin $\textit{et al.}$ \cite{Mezhov1975} carried out a comprehensive experimental and numerical study to observe the transition between Poiseuille flow and Knudsen flow in Bi crystal at very low temperatures ($<$ $2 K$). At temperatures below 1.3 $K$ and with diameter below 0.5 cm, neglecting the impurity and defect scattering processes, the Knudsen minimum was found \cite{Mezhov1975} to exist at temperature $\sim$ 1 $K$, represented through the minimum of the size variation of the effective mean free path ($l_{eff}$). Solid He$^4$ \cite{Armstrong1979} was also found to feature Knudsen minimum in the mean free path at $T \approx 0.25 K$. Guo and Wang \cite{Guo2017} developed a numerical method to solve Boltzmann equation using Callaway's \cite{Callaway1959} dual relaxation model to study heat transport in two-dimensional materials and found the existence of Knudsen minimum in the graphene ribbon. The minimum was shown \cite{Guo2017} to persist at low temperatures when the average normal scattering rate seems to strongly dominate (around 100 times stronger) than the resistive scattering rates, realized via the width variation of the nondimensional heat flow rate. Ding $\textit{et al.}$ \cite{Ding2018} extensively used first-principles calculations to obtain the exact solution of BTE for graphite and surprisingly identified Knudsen minimum at significantly higher temperature ($\sim$ 90 K) opening up the possibilities to discover Knudsen minimum in other 3D materials. Soon after, the experimental study by Martelli $\textit{et al.}$ \cite{MartelliStrontium2018} identified the Knudsen minimum in SrTiO$_3$ via the minimum in the temperature variation of the effective mean free path. Experimental efforts by Machida $\textit{et al.}$ \cite{Machida2018P} explored the thickness dependence of the Knudsen minimum in the temperature variation of the effective mean free path along with the prominent Poiseuille flow characteristics in black P. They found that increasing the sample thickness gradually shifts the Knudsen minimum to the lower temperatures. Increasing thickness helps larger pool of phonons to undergo normal scattering which essentially triggers more diffuse boundary scattering at the onset of ballistic regime, leading to this trend. Another study \cite{Machida309} on thin graphite sample was also found to exhibit Knudsen minimum around 10 K. A multiscale computational protocol \cite{Luo2019} for solving the transient BTE using Callaway's dual relaxation model was employed to investigate the second sound in graphene and identified the sample width (2 $\mu$m) where the Knudsen minimum was observed at 40 K via the sample width variation of the nondimensional thermal conductivity. Using deviational Monte Carlo method with first-principles calculations, Li $\textit{et al.}$ \cite{Li2019} numerically solved BTE to investigate the crossover between different heat conduction regimes in suspended graphene. Nondimensional thermal conductivity was realized via $\kappa'$ (= $\kappa/W$) which is similar to the original dimensionless thermal conductivity $\kappa^{*}$ = $\kappa T_0/(C\overline{v}W)$, where $C$, $\overline{v}$ and $W$ represent energy density, average group velocity and the width of the sample, respectively \cite{Li2019}. By decomposing $\kappa'$ into ballistic and scattered contributions, Knudsen minimum was observed to be present at 100 K at width $\approx$ 0.7 $\mu$m for suspended graphene \cite{Li2019}. In a recent $\textit{ab initio}$ study \cite{kanka3}, three dimensional crystalline GeTe was found to exhibit a shallow Knudsen minimum like feature at low temperature marking the onset of the phonon hydrodynamics. Figure \ref{fig:knudsen_minimum} outlines various works with a strong presence of Knudsen minimum in various solids.

\section{Phonon hydrodynamics from material science perspective}{\label{sec:5}}

\begin{figure}[H]
    \centering
\includegraphics[width=1.0\textwidth]{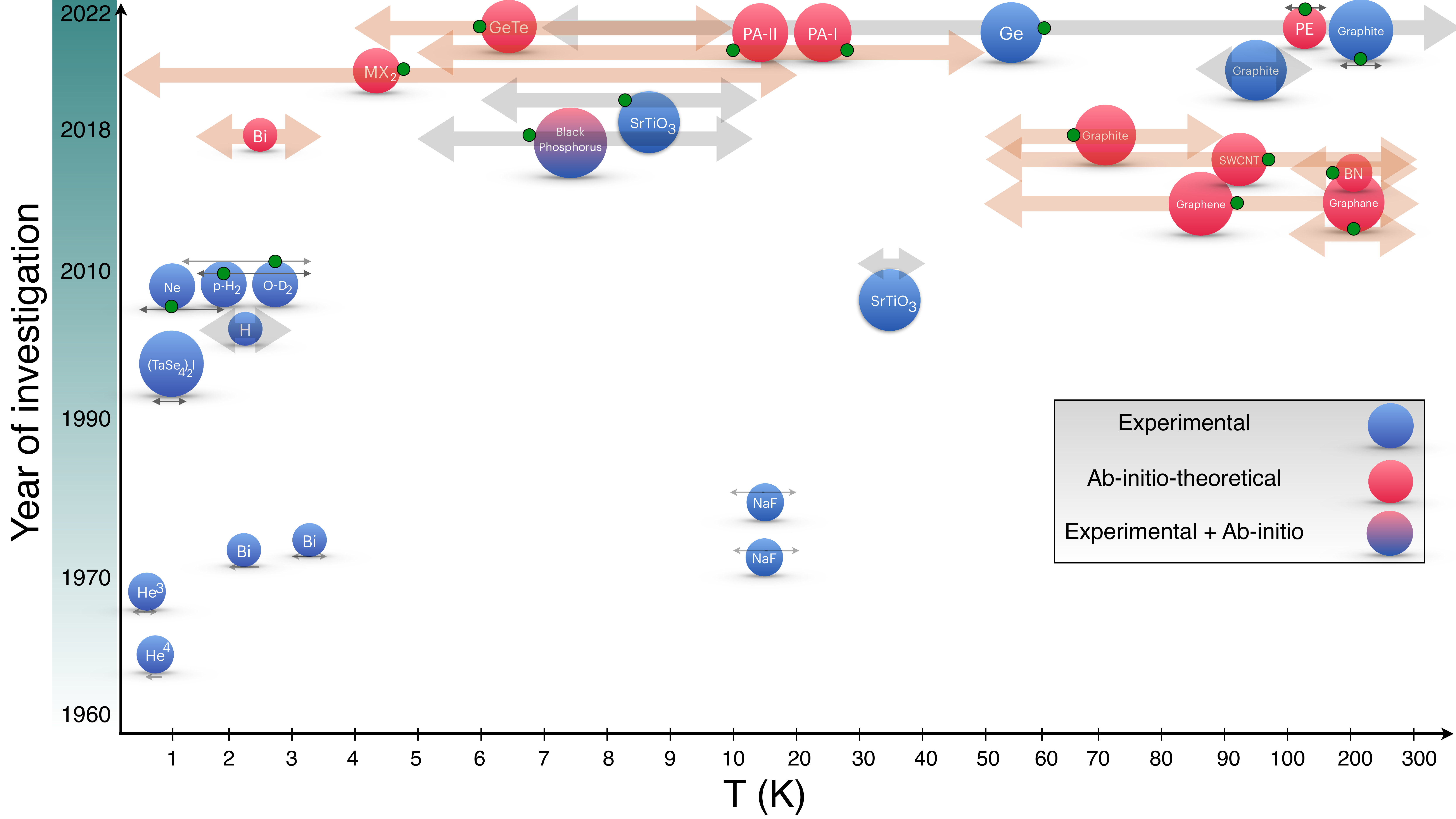}
    \caption{A comprehensive up-to-date review of various 3D, 2D and 1D materials that have been found to feature phonon hydrodynamic signatures via various experimental, numerical and theoretical methods. The materials, supporting phonon hydrodynamics, are represented in a parameter space of temperature and the year of their investigation. For each of the materials, the corresponding temperature windows for featuring phonon hydrodynamics, has been defined via the double headed arrows. Also, different investigations related to the experimental and theoretical-numerical methods have been distinguished. As it is evident from the gradual growing resources of advanced computational and experimental techniques, explorations regarding phonon hydrodynamics are more frequent after 2010 compared to the era before it. For different materials at similar temperature range or similar year of study, green dots are marked for clarity to guide the eye to distinguish the corresponding temperature window for specific materials. Some of the uncommon symbols for the materials are given below: p-H$_2$: parahydrogen \cite{Khodusov2009}; o-D$_2$: orthodeuterium \cite{Khodusov2009}; MX$_2$: (M= Mo, W; X=S, Se) \cite{Torres_2019}; PA-I: polyacene \cite{ZhangPolymer2020}; PA-II: polyacetylene \cite{ZhangPolymer2020}; PE: polyethylene \cite{ZhangPolymer2020}.}
    \label{fig:overview}
\end{figure}

While phenomenological understanding is crucial from the physicist's point of view, a material science perspective is also equally important in the current age of advanced material science, process engineering, metallurgy and applied physics. Physical understanding of the hydrodynamic phenomena in solids eventually should lead to harnessing the suitable properties to produce materials of interests which can solve various heat conduction related problems and open up new areas of scientific explorations. Though some of the reviews \cite{Lindsay2019, GangChenReview2021} on the subject briefly described the materials that feature phonon hydrodynamics, a thorough up-to-date account of the materials of interest seems essential. Figure \ref{fig:overview} summarizes an up-to-date account of the materials that possess phonon hydrodynamics with corresponding temperature window of occurrence and their year of study, investigated via experiments and advanced theoretical/computation techniques.

\subsection{\textbf{3D materials}}

As discussed in earlier studies \cite{BeckReview1974, Lindsay2019, GangChenReview2021, Leebookchapter2020}, it demands a stringent set of rules for the materials to qualify for persisting and exhibiting phonon hydrodynamics phenomena. Extremely low, often cryogenic temperature to switch on enough normal scattering or appropriate size of the sample to satisfy GK condition, make it difficult for materials to become a suitable candidate for phonon hydrodynamic phenomena e.g. second sound, Poiseuille flow, Knudsen minimum etc. Furthermore, isotopic purity, vacancies in the crystal structure, structural instability are also found to influence the operational regime of phonon hydrodynamics. No wonder, only few materials had been found over the years to substantially exhibit phonon hydrodynamics. Initially, works of Tisza \cite{Tisza1938} and Landau \cite{Landau1941, Landau1947} and later the experimental works by Peshkov \cite{Peshkov1944, Peshkov1948} on the second sound wave propagation in liquid He evoked curiosity about the existence of phonon hydrodynamics in the solids. Using heat-pulse experiments phonon hydrodynamics regime was obtained for solid He$^4$ \cite{Ackerman1966solidhelium, Mezhov-Deglin1965, Hogan1969}. For solid He$^4$, Ackerman $\textit{et al.}$ \cite{Ackerman1966solidhelium} found the operational temperature regime to be extremely low (below 0.7 K). Similar heat-pulse propagation experiments were also carried out \cite{Thomlinson1969} for solid bcc He$^3$ and the signatures of phonon hydrodynamics were found to be present at $T$ $\leq$ 0.58 K \cite{Thomlinson1969, Ackerman1969solidhe3}. Both heat-pulse \cite{Jackson1970NaF} as well as light scattering experiments \cite{Pohl1976NaF} showed consistent phonon hydrodynamic behavior in the temperature range $10$ K $<$ $T$ $<$ $20$ K in the crystalline NaF. A similar fluoride material, LiF was also investigated by Bausch $\textit{et al.}$ \cite{Bausch1972} in this context. Dislocation density and thickness of these fluoride crystals were found to act as critical factors, making it difficult to realize phonon hydrodynamic modes in these samples \cite{Bausch1972}. Bi with sufficient isotopic purity was experimentally observed \cite{Narayanmurti1972Bi, Kopylov1971, Kopylov1973} to feature phonon hydrodynamics at low temperature regime ($3$ K $<$ $T$ $<$ $3.5$ K) which had been verified only few years back by Markov $\textit{et al.}$ \cite{Bitheory2018} via the first-principles calculations coupled with variational solution of the LBTE. Both doped and undoped SrTiO$_3$ had been experimentally investigated \cite{MartelliStrontium2018, Koreeda2007SrTio3} in the context of phonon hydrodynamics and Poiseuille regime with faster than $T^3$ scaling was observed for thermal conductivity in the temperature range $6$ K $<$ $T$ $<$ $13$ K for undoped SrTiO$_3$. Second-principles density functional theory with full LBTE solutions by Torres $\textit{et al.}$ \cite{Torres2019} confirmed the hydrodynamic phonon transport in SrTiO$_3$ with smaller characteristic size of the sample (4 $\mu$m). Contrary to the conventional solids, SrTiO$_3$ hosts soft optical modes even at low temperature \cite{MartelliStrontium2018}. Therefore, unlike the solid He crystals, where chemical purity is crucial, the large contribution of the three phonon phase space, particularly normal scattering events, helps SrTiO$_3$ to exhibit phonon hydrodynamics. In another approach, the quasielastic light scattering experiment \cite{Koreeda2009SrTio3} on SrTiO$_3$ in the low frequency range revealed an anomalously broad doublet structure in the light scattering spectrum in the temperature interval of $30$ K $<$ $T$ $<$ $40$ K \cite{Koreeda2009SrTio3, Hehlen1995SrTio3, Koreeda2007SrTio3}. This doublet was argued as the signature of second sound whose presence had been a subject of debate \cite{SCOTT2000185}. Machida $\textit{et al.}$ \cite{Machida2018P} experimentally observed phonon hydrodynamics in black P in the temperature range $5$ K $<$ $T$ $<$ $12$ K utilizing the Poiseuille flow characteristics in the temperature variation of the effective mean free path. Recently, thermal grating measurement \cite{Huberman2019graphite}, thermal conductivity assessment \cite{Machida309} as well as first-principles simulations \cite{Ding2018} reported phonon hydrodynamic features in graphite at substantially high temperature. While experiments recorded the second sound propagation between $85$ K and $125$ K \cite{Huberman2019graphite}, simulations revealed a corresponding temperature window of $50$ K $<$ $T$ $<$ $90$ K \cite{Ding2018}. However, very recent findings by Ding $\textit{et al.}$ \cite{Ding2022} recorded second sound in graphite at even higher temperatures (at 200 K and at 225 K) using sub-picosecond transient grating method supported by $\textit{ab initio}$ simulations. Recently, bulk Ge \cite{Beardo2021Ge} has been shown to exhibit phonon hydrodynamics in a wide temperature range ($7$ K $<$ $T$ $<$ $300$ K) using a rapidly varying temperature field and observing the phase lag of the thermal response of the material supported by $\textit{ab initio}$ and nonequilibrium MD studies. Some of the other materials that feature phonon hydrodynamics include solid H \cite{Zholonko2006}, orthodeuterium, parahydrogen quantum crystals, neon crystals \cite{Khodusov2009} and crystalline polymers (Polyacene, Polyacetylene, Polyethylene) \cite{ZhangPolymer2020}.

\subsection{\textbf{2D materials}}

2D materials are perhaps the most studied materials in recent times in the context of the phonon hydrodynamics as hydrodynamic phonon transport manifests striking features(e.g. high thermal conductivity and wider temperature window for hydrodynamics in graphene) compared to the 3D materials, marking a significant influence on the technological and heat transfer applications \cite{GuRevModPhys, Cepellotti2015, Scuracchio2019numerical, Broido}. Especially, the second sound phenomena (fast thermal conduction with negligible damping) can be exploited to use graphene as potential thermal signal transmitters \cite{Broido}. Extremely high thermal conductivity ($\approx$ 4000 W/mK for suspended graphene at room temperature \cite{Li2018}) and substantial presence of phonon hydrodynamics in 2D materials stem from their unusual anharmonic interaction associated with the ZA (flexural acoustic) modes. Unlike its 3D counterparts, 2D materials possess two different dispersion relations, linear and quadratic, corresponding to the in-plane and out-of-plane (flexural) displacements respectively. These ZA modes were found to contribute significantly to the heat conduction in suspended graphene giving rise to the extremely high thermal conductivity in graphene \cite{Seol2010, Lindsay2010}. Michel $\textit{et al.}$ \cite{Michel2015} found that in a broad temperature range ZA modes are less affected by Umklapp scattering compared to the in-plane modes and therefore supported the realization of large intrinsic thermal conductivity of graphene. Because of the serious contribution of the out-of-plane flexural modes and its quadratic dispersion, 3D Debye model is insufficient to precisely describe 2D systems like graphene, BN etc. Therefore, different state-of-the-art numerical and theoretical strategies were adopted for 2D materials to accurately predict thermal transport in the presence of phonon hydrodynamics. Recently, Shang $\textit{et al.}$ \cite{ShangSrep2020} derived 2D GK equation by taking the quadratic dispersion into account and found that the second sound speed in graphene varies with temperature. This result is very different from the regular Debye model for both 3D and 2D systems where second sound velocities were found to be temperature independent with values $v_{I}/\sqrt{3}$ and $v_{I}/\sqrt{2}$ respectively \cite{ShangSrep2020, YuReview2021}, $v_I$ being the sound speed. $\textit{Ab initio}$ methods were extensively used in association with iterative \cite{Broido} or variational approach \cite{Cepellotti2015} to solve LBTE for 2D materials and revealed an extremely strong N scattering events at a wide range of temperatures up to room temperature \cite{Cepellotti2015} as well as large density-of-states of the long-wavelength ZA phonons \cite{Broido}. Scuracchio $\textit{et al.}$ \cite{Scuracchio2019numerical} derived a coupled integrodifferential equations for acoustic sound waves and phonon density fluctuations and eventually derived hydrodynamic equations for 2D two-dimensional (2D) crystals. Using Callaway's dual relaxation model, Guo and Wang \cite{Guo2017} developed a discrete-ordinate-method to study heat transport in 2D materials. Li $\textit{et al.}$ \cite{Li2018, Li2019} studied the hydrodynamic phonon transport in suspended graphene using deviational Monte Carlo scheme coupled with $\textit{ab initio}$ method to obtain the scattering matrix. $\textit{Ab initio}$ LBTE studies \cite{Cepellotti2015, Broido} reveal a wide temperature window for featuring phonon hydrodynamics for graphene ($\approx$ 50 K - 300 K), BN ($\approx$ 100 K- 300 K) and graphane ($\approx$ 100 K- 300 K) employing GK conditions related to the analysis of scattering rate hierarchy. Torres $\textit{et al.}$ \cite{Torres_2019} investigated low-thermal conductivity 2D metal dichalcogenide materials (MoS$_2$, MoSe$_2$, WS$_2$, WSe$_2$) using first-principles method with Kinetic Collective model adopted from GK hydrodynamic equation and found phonon hydrodynamic window below 20 K, using the nonlocal length assessment, present in the GK type hydrodynamic equation. Due to isotopic abundance, the difference between N and resistive scattering rates were found to be lower for metal dichalcogenides compared to the graphene \cite{Torres_2019}. Readers are also recommended recent perspective articles \cite{YuReview2021, Liureview21} which cover the phonon hydrodynamics and its implications in 2D materials.

\subsection{\textbf{1D materials}}

Apart from graphene and other 2D materials, 1D materials like single-walled carbon nanotubes (SWCNT) also possess high thermal conductivity and therefore had been envisaged as a potential candidate to feature phonon hydrodynamics in the literature. This high thermal conductivity of SWCNT was found to be associated with high Debye temperature which facilitates large group velocities of acoustic phonons and feeble Umklapp scattering at room temperature \cite{Lee2017}. Osman $\textit{et al.}$ \cite{Osman2005MD} investigated the heat pulse propagation in armchair (5,5) and zig-zag (10,0) and (7,0) SWCNT using MD simulations and found a significant contribution of the second sound waves to carry the energy of the heat pulse. Lee $\textit{et al.}$ \cite{Lee2017} employed lattice dynamics calculations to solve BTE (Peierls-Boltzmann transport equation) to study the hydrodynamic phonon drift and second sound propagation in a (20,20) SWCNT. A wider temperature window was observed to feature phonon hydrodynamics (50 K $<$ $T$ $<$ 300 K) with a considerable amount of total heat was shown to be transferred by the drifting phonons ($\approx$ 70 $\%$ and 90 $\%$ at 300 K and 100 K respectively). Thermal conductivity measurements of quasi-one-dimensional (TaSe$_4$)$_2$I single crystals \cite{Smontara1996} revealed sharp peaks around 1 K, which had been argued as a manifestation of the phonon Poiseuille flow.

\subsection{\textbf{Low thermal conductivity materials}}

Naturally, most of the research related to phonon hydrodynamics revolved around 2D and high thermal conductivity materials as these materials are more prone to feature second sound, Poiseuille flow and therefore emerge as natural choices for their manipulation for industrial applications. Torres $\textit{et al.}$ \cite{Torres_2019} numerically explored the phonon hydrodynamics for comparatively low thermal conductivity metal dichalcogenide materials and found that at low temperatures they exhibit phonon hydrodynamic features. However, the study focused on the single layer transition metal dichalcogenides and explored the scenario where the materials possess low thermal conductivity despite being 2D systems. In a recent series of investigations \cite{kanka2, kanka3}, phonon hydrodynamics had been put under inspection for even lower thermal conductivity, 3D chalcogenide phase change materials. These investigations, focused on GeTe, a low thermal conductivity chalcogenide with phase change memory applications, revealed an unusual presence of phonon hydrodynamics at low temperatures. The $\textit{ab initio}$ study \cite{kanka2} coupled with complete solution of LBTE using direct method \cite{chaput} as well as Kinetic Collective model \cite{Torres2017KCM, alvarez2018thermalbook} demonstrated that phonon hydrodynamics criteria are met at low temperature in GeTe provided a favourable condition relating larger grain size and lower vacancy scattering events are satisfied. Isotope scattering rate emerged as an identifier to distinguish the presence and absence of the hydrodynamic window via the conditions $\tau_{I}^{-1} (\omega)$ $>$ $\tau_{V}^{-1} (\omega)$ and $\tau_{I}^{-1} (\omega)$ $<$ $\tau_{V}^{-1} (\omega)$  respectively, where $\tau_{I}^{-1} (\omega)$ and $\tau_{V}^{-1} (\omega)$ denote isotope scattering rate and phonon-vacancy scattering rate respectively, as a function of phonon frequency. Further, the hydrodynamic window for low $\kappa$ material GeTe was found to be very fragile and sensitive towards the competition between 

\noindent temperature and grain size. Systematic investigation of thermal transport as a function of characteristic size \cite{kanka3} revealed the complete hydrodynamic window in temperature-grain-size plane via Knudsen number and average scattering rate analysis using GK conditions. Moreover, the scaling of thermal conductivity ($\kappa$) with characteristic length ($L$) was found to dictate the existence of the Knudsen minimumlike prominent hydrodynamic feature \cite{kanka3}. Between ballistic (linear scaling of $\kappa$ with $L$) and diffusive ($L$ independent $\kappa$) thermal transport regime, a superlinear scaling in the intermediate $L$ regime was found assisting a Knudsen minimumlike hydrodynamic feature at a particular temperature \cite{kanka3}. On the contrary, sublinear scaling leads to weak phonon hydrodynamics. The notable visibility of collective phonon transport in GeTe was found to be controlled by a ratio of average normal and resistive scattering rates. The quest of phonon hydrodynamics in low $\kappa$, 3D chalcogenide GeTe answered some fundamental issues: (a) It was understood that even for a very low thermal conductivity materials phonon hydrodynamics can be observed if peculiar conditions relating grain size, temperature and vacancy scattering rates are satisfied. (b) Identifying the controlling parameters can help understanding the generic behavior of the phonon
hydrodynamics in low thermal-conductivity materials.

\section{The controlling parameters of phonon hydrodynamics}{\label{sec:6}}

\begin{figure}[H]
    \centering
\includegraphics[width=0.9\textwidth]{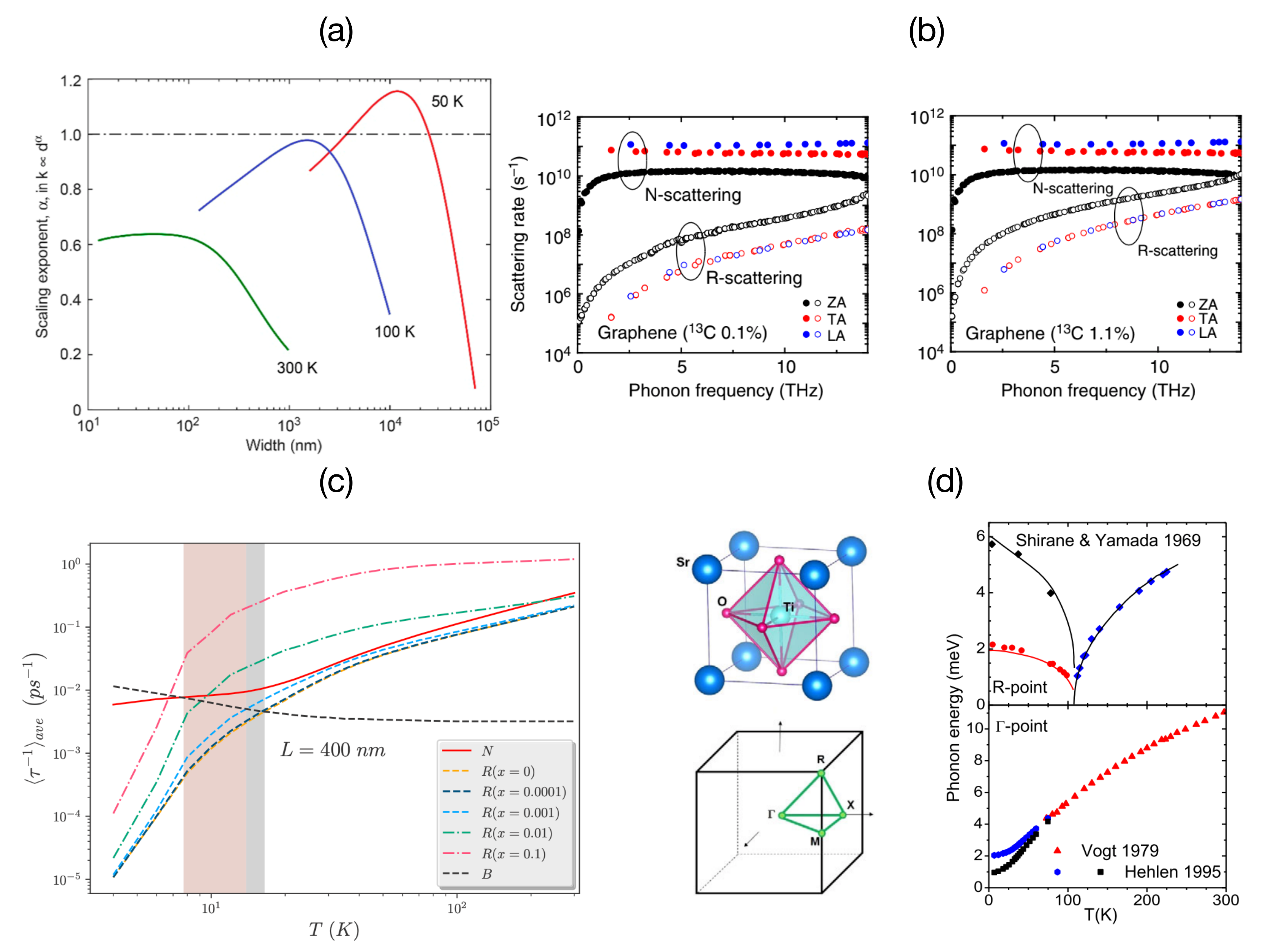}
    \caption{\footnotesize Various influencing factors for featuring phonon hydrodynamics in solids. (a) Effect of sample width on featuring phonon hydrodynamics, realized using the scaling exponent ($\alpha$ = ${\partial ln(\kappa)}/{\partial ln(d)}$) of the width ($d$) variation of thermal conductivity of graphite ribbon. At 50 K, $\alpha$ $>$ 1 which is a marker of having phonon hydrodynamics, contrary to the $T$ = 100 K and 300 K. Reprinted (figure) with permission from Ref. \cite{Ding2018}. Copyright (2018) American Chemical Society. (b) Effect of isotopes on phonon hydrodynamics realized in $\textit{ab initio}$ simulations of graphene with two different isotope contents (0.1 $\%$ and 1.1 $\%$). R scattering is found to increase for higher isotopically enriched graphene sample, lowering the gap between N and R scattering. Reprinted by permission from Springer Nature Customer Service Centre GmbH: [Springer Nature] [Nature Communications] \cite{Broido} (2015). (c) Effect of vacancies on phonon hydrodynamics: first-principles simulations of crystalline GeTe of fixed grain-size showed a cut-off vacancy density ($x$ = 0.001 $\%$) in the crystal above which hydrodynamic window vanishes and below which the hydrodynamic window opens up. Reprinted (figure) with permission from Ref.  \cite{kanka2}. Copyright (2020) American Physical Society. (d) Effect of structural instability on phonon hydrodynamics: Crystal structure of SrTiO$_3$ possesses two soft modes and the temperature dependence of these two soft modes are shown. These low frequency soft modes can interact with acoustic modes to give a strong anharmonicity favourable for phonon hydrodynamics. Reprinted (figure) with permission from Ref. \cite{MartelliStrontium2018}. Copyright (2018) American Physical Society.}
    \label{fig:controlling_parameter}
\end{figure}

The earlier discussions in this review lead to the central idea that the existence, persistence and prominence of the phonon hydrodynamics are mostly dependent on the strong presence of the N scattering events compared to the other resistive phonon scattering processes. Therefore, manipulating phonon hydrodynamics boils down to the situation where the related parameters are tuned in such a way so that the N scattering processes show prominence. After sufficient occurrences of the N scattering events, the phonon distribution changes from Bose-Einstein to a displaced Bose-Einstein distribution with a drift velocity causing a net flow of phonons in the direction of heat conduction. It has been thoroughly discussed in literature \cite{Leebookchapter2020, GangChenbook, kaviany_2014} that N scattering cannot give rise to thermal resistance on its own. Either intrinsic resistive scattering events (Umklapp, isotope or vacancy scattering) or the characteristic length of the sample (related to the grain-boundary scattering) induces thermal resistance to yield a finite thermal conductivity of a material. In this section, we discuss about these parameters and how they influence the phonon hydrodynamics. Figure \ref{fig:controlling_parameter} demonstrates a summary of various factors studied in the literature that influence the occurrence of the phonon hydrodynamics in a material.

\subsection{Effect of thickness and characteristic length:}

Once N scattering alters the phonon population from following the Bose-Einstein to a displaced Bose-Einstein distribution, a drift motion of phonons emerges that drives the phonon flow in the heat flow direction. Once set up, this drift motion of phonons can even sustain without any temperature gradient \cite{Leebookchapter2020, LeeMC2019}. For an infinitely large material, unless Umklapp scattering is considered, N scattering can lead to infinite thermal conductivity. This can be understood by the study of Lindsay $\textit{et al.}$ \cite{Lindsay2009} on the lattice thermal conductivity of SWCNT. Though phonons with small wave vectors are seen to satisfy mostly the occurrence of the N scattering, they can scatter with phonons of long wave vectors through N scattering and facilitate U scattering and thus switching on the thermal resistance \cite{Leebookchapter2020}. It was shown by Lindsay $\textit{et al.}$ \cite{Lindsay2009} that the scattering between acoustic and optical phonons are necessary to incorporate Umklapp scattering and therefore thermal resistance which in turn significantly reduce the room temperature thermal conductivity of SWCNT.

A material with infinite length but finite width (width $\gg$ mean free path of N scattering) brings about the diffuse boundary scattering due to the grain boundaries and therefore induces thermal resistance. As noted in earlier section in the context of Poiseuille flow of phonons, drift velocity near the boundaries are heavily reduced due to diffuse boundary scattering of phonons compared to that of the centre of the width of the sample. This drift velocity gradient across the width (perpendicular to the heat flow direction) assists in imparting the momentum from the centre to the boundary. However, N scattering events impede this momentum transfer via phonon hydrodynamic viscosity. This viscous damping decreases with the quadratic power of width ($W$) as was realized via momentum balance equation of phonons \cite{Li2018, Leebookchapter2020} and as a result, thermal conductivity in the phonon hydrodynamic regime increases superlinearly with the width of the sample. Here we note that in a situation of negligible presence of Umklapp scattering and zero drift velocity at the boundary, thermal conductivity actually scales as $W^2$. However, the unavoidable presence of Umklapp scattering in the real samples forces the exponent to take the value $\alpha$, where 1 $<$ $\alpha$ $<$ 2 \cite{Ding2018, Li2018, kanka3}. This superlinear width dependency of $\kappa$ is strikingly distinct from the ballistic and the diffusive thermal transport regimes where $\kappa$ varies linearly and stays constant with the width of the sample, respectively \cite{Broido, Ding2018, kanka3}.

Lee $\textit{et al.}$ \cite{LeeMC2019} explored another scenario with graphitic samples having infinite width but finite length between hot and cold reservoirs in the heat flow direction, resembling cross-plane heat flow in thin-film using deviation Monte Carlo scheme. It was shown that even without Umklapp scattering, thermal resistance can be facilitated via the N scattering when non-collective flow of phonons transformed into collective flow of phonons due to the finite length of the sample (larger than the N scattering mean free path). This resistance caused by N scattering with finite length was found to be dictated by the shape of the phonon dispersion and more specifically nonlinear phonon dispersion for the graphitic materials \cite{LeeMC2019}. This nonlinear phonon dispersion causes significant entropy generation compared to that of the Debye model and gives rise to thermal resistance in graphitic materials \cite{LeeMC2019, Leebookchapter2020}. Very recently, Nie and Cao \cite{Ben2022} explored the boundary and interfacial thermal behavior in 2D systems in the context of phonon hydrodynamics, using Monte Carlo simulation algorithm described in \cite{NieMC2020}. Two cases had been studied: (a) a nanofilm with finite length and infinite width, similar to that of Lee $\textit{et al.}$ \cite{LeeMC2019}, and (b) two nanofilms of the same material connected to each other, making an interface. The interfacial behaviors were seen as the sum of the interactions in two isolated nanofilms with the interface effects, reasonably supported via numerical simulations.

\subsection{Effect of isotopes and vacancies:}

Real samples normally exhibit defects, isotopes and impurities during the crystallization process which affect the coherent phonon flow significantly. Isotope and vacancies in a sample directly reduce the probability of occurring hydrodynamic phonon flow and shrink the phonon hydrodynamic window in a sample as they hinder the N scattering events resistively. Using lattice dynamical model and second-order perturbation theory, Tamura \cite{Tamura} derived the scattering rate of phonons by randomly distributed isotopes ($\tau_{I}^{-1}$) in a material as

\begin{equation}
    \frac{1}{\tau_{\lambda}^{I}(\omega)} = \frac{\pi \omega_{\lambda}^{2}}{2N}\sum_{\lambda'} \delta\left(\omega - \omega'_{\lambda} \right) \sum_{k} g_{k}|\sum_{\alpha}\textbf{W}_{\alpha}\left(k,\lambda \right)\textbf{W}_{\alpha}^{*}\left(k,\lambda \right)| ^{2} 
\end{equation}
Here, $g_k$ is the mass variance parameter, defined as 
\begin{equation}
    g_{k} = \sum_{i} f_{i} \left( 1 - \frac{m_{ik}}{\overline{m}_k}\right)^{2}
\end{equation}
where $f_i$ is the mole fraction, $m_{ik}$ denotes the relative atomic mass of $i$th isotope, $\overline{m}_k$ is the average mass = $\sum_{i} f_{i} m_{ik}$, and $\textbf{W}$ is a polarization vector. The similar idea of mass variance due to point defect was employed by Ratsifaritana and Kelemens \cite{ratsifaritana} using a perturbation technique to derive phonon scattering rates by vacancy defects by estimating missing mass and missing linkage between the masses. The phonon-vacancy scattering rate is realized as \cite{ratsifaritana}
\begin{equation} {\label{eqvacancy}}
    \frac{1}{\tau_{V}(\omega)} = x \left(\frac{\Delta M}{M}\right)^{2}\frac{\pi}{2}\frac{\omega^{2}g(\omega)}{G'}
\end{equation}
where, $x$ is the vacancy concentration, $G'$ defines the number of atoms in the crystal, and $g(\omega)$ denotes the phonon density of states. Realizing vacancies as isotope impurity, 
Ratsifaritana $\textit{et al.}$ \cite{ratsifaritana} estimated mass change $\Delta M$ = 3 M, where M is the mass of the removed atom. This comes from the fact that a vacancy is equivalent to omitting one atom from the material and all the linkages between the removed atom and its neighbor. As every linkage is connected to two atoms, removing one atom with two linkages costs the removal of another two atoms, associating a mass change of $\Delta M$ = 3 M.

Both isotope and vacancy scattering rates are found to increase with phonon frequency. While considering the average scattering rates for isotopes, the weighted average of modal heat capacity makes it prominent at low temperatures and thus increases the resistive scattering rates. Moreover, the phonon scattering rates by vacancies (Eq.\ref{eqvacancy}) are found to vary linearly with the vacancy density, indicating its larger contribution due to increased vacancies in the material. These resistive scattering rates add up with the already intrinsically present Umklapp scattering and the the total resistive phonon lifetime ($\tau_R$) is realized using Matthiessen's rule as \cite{kaviany_2014} 
\begin{equation}{\label{matthiessen}}
    \frac{1}{\tau_{R}} = \frac{1}{\tau_U} + \frac{1}{\tau_I} + \frac{1}{\tau_V} + \frac{1}{\tau_B}
\end{equation}
and hinder the collective phonon flow driven by the N scattering events, making the hydrodynamic window more fragile and narrower. Here $\tau_U$, $\tau_I$, $\tau_V$ and $\tau_B$ are phonon lifetimes corresponding to the Umklapp, isotope, vacancy and boundary scattering respectively.  

Solid helium \cite{Ackerman1966solidhelium}, NaF \cite{Jackson1970NaF} are some of the materials that are isotopically pure to visualize the signatures of phonon hydrodynamics. Though chemical purity of Bi is lesser compared to the solid helium, it exhibits sufficient isotopic purity to feature phonon hydrodynamics \cite{Narayanmurti1972Bi, Bitheory2018}. Effects of vacancies were also found to be extremely sensitive towards

\noindent opening of the hydrodynamic window as shown for GeTe \cite{kanka2}. Further, average isotope scattering rate was found \cite{kanka2} to act as a marker in choosing appropriate vacancy density to enable a hydrodynamic window.

\subsection{Effect of structural instability:}

Sometimes, even isotopically not so pure substances can still possess the features of phonon hydrodynamics thanks to its structural instability which acts in enhancing the anharmonicity (or large three phonon phase space) and therefore the N scattering processes. SrTiO$_3$ is a representative case where anharmonicity driven N scattering was found to play a crucial role in featuring phonon hydrodynamics as studied in several experimental and theoretical studies \cite{STO, MartelliStrontium2018, Hehlen1995SrTio3, Koreeda2007SrTio3, Koreeda2009SrTio3}. SrTiO$_3$ falls under the category of displacive ferroelectric material which consists of a TiO$_6$ octahedra and has strontium atoms at its vertices in its cubic elementary cell and was found to exhibit two soft modes located at R point and zone center respectively \cite{MartelliStrontium2018}. Gurevich $\textit{et al.}$ \cite{STO} theoretically indicated the possibilities of ferroelectric materials to show phonon hydrodynamics due to the presence of the soft modes in this class of materials. Unlike regular solids, a displacive ferroelectric hosts at least one optical mode (soft mode) at the center of the Brillouin zone whose frequency is anomalously low \cite{STO}. N scattering events can be observed to be frequent due to the strong interactions between these low frequency (long wavelength) soft modes (optical phonons) and the acoustic phonons. It was proposed that the second sound waves, produced in such ferroelectrics with a wider frequency interval (of the order of 10 GHz), can be experimentally observed via the ordinary light scattering experiments. This was validated by Martelli $\textit{et al.}$ \cite{MartelliStrontium2018} via the thermal conductivity measurements of SrTiO$_3$.

We also note here the property of exhibiting flexural acoustic (ZA) modes (for graphene like 2D materials) as an important controlling parameter for realizing phonon hydrodynamics which had been discussed in the earlier section dedicated to the 2D materials.

\section{Summary and Outlook}{\label{sec:7}}

In this review, phonon hydrodynamics in crystalline materials has been discussed from both phenomenological and material science perspectives. Starting from the microscopic understanding of the phonon scattering, the subject has been approached via theoretical, experimental and numerical explorations. In all these different methods, a chronological, state-of-the-art account of the subject starting from employing kinetic theory to the advanced relaxon approach is described. Three of the most prominent features of the phonon hydrodynamics: second sound, Poiseuille flow and Knudsen minimum have been chosen and thoroughly represented from the phenomenological perspective along with their distinct methods of investigations. The criteria for occurring phonon hydrodynamics via these three realizable phenomena are also described in detail pertaining to theoretical, numerical and experimental results. The advanced numerical methods involving $\textit{ab initio}$ techniques greatly helped in improving the accuracy of the solution of the Linearized Boltzmann transport equation which is significantly crucial for the identification of the signatures of phonon hydrodynamics via thermal conductivity calculations. Apart from the physical phenomena based approach, the subject has also been discussed from the material science perspective, which is equally important if not more. This perspective seems decisive in carrying out applications related to phonon hydrodynamics. In this context, a thorough, up to date review of the materials, ranging from three, two and one dimensional systems, has been carried out which exhibit phonon hydrodynamics as was realized via theoretical or experimental observations. Though 2D materials emerge as the most promising candidates to harness several crucial applications (e.g. graphene as efficient thermal rectifier and thermal signal transmitter \cite{Broido}) due to their strong N scattering features due to the anharmonicity caused by flexural modes, 3D materials with both high and low thermal conductivity cases are discussed. Chalcogenide low thermal conductivity material GeTe, used mostly as phase change memory devices, has newly been realized \cite{kanka2, kanka3} to feature phonon hydrodynamics if the controlling parameters are adjusted carefully. Though the low conductivity and low temperature occurrence of phonon hydrodynamics of this class of materials limit them from some technological applications, nevertheless this helps to predict more accurate thermal conductivity beyond the relaxation time applications. Moreover, this raises the question and opens up the avenues to understand the generic behavior and manifestations of phonon hydrodynamics in a better way in low thermal conductivity materials. With the help of advanced first-principles techniques, all these studies indicate the subtle presence of phonon hydrodynamics in a variety of materials. To dig up this subtle existence and consequences of phonon hydrodynamics, it is imperative to study the controlling parameters of phonon hydrodynamics. Therefore, we review the effect of characteristic size, defects and instabilities on the prominent occurrence of phonon hydrodynamics. This panoramic exploration of phonon hydrodynamics will enable us understanding the captivating physics behind these features as well as methods and ways to harness engineering applications (thermal rectification, thermal dissipator etc) for non metallic solids in general. We note here two instances to relate two important phonon hydrodynamic features, second sound and Poiseuille flow with two potential applications, namely thermal rectification and thermal interconnect applications, respectively. Thermal rectification is a process which allows heat transfer in one direction but block the other \cite{rectifier}. This feature has several important implications in the thermal management of micro and nanoelectronics concerning thermal diode, thermal logic gates etc \cite{Physrep2015Guo}. Poiseuille flow of phonons can provide a better understanding of the thermal rectification process. Moreover, the effect of diffusive boundary scattering on the drift motion of phonons can be useful to study the rectification efficiency of nanoscaled devices with certain boundary roughness \cite{Physrep2015Guo, Broido}. The development of micro and nanoelectronics also compels more thermally and electrically efficient, smaller interconnects in the integrated circuits and therefore graphene appears to be a better alternative to the already existing Cu interconnects \cite{sonInterconnect2021, behnamInterconnect2012}. However, there are issues of concern related to the high resistance of single layer graphene nanoribbons for this application \cite{dasInterconnect2018, muraliInterconnect2009}. The signature of second sound, a fast thermal transport with negligible damping, can help improving the understanding of graphene as thermal interconnects \cite{Broido}. Thus, exploiting and manipulating second sound phenomena of graphene can further help improving the efficiency of the graphene thermal interconnects.

\begin{figure}[H]
    \centering
\includegraphics[width=1.0\textwidth]{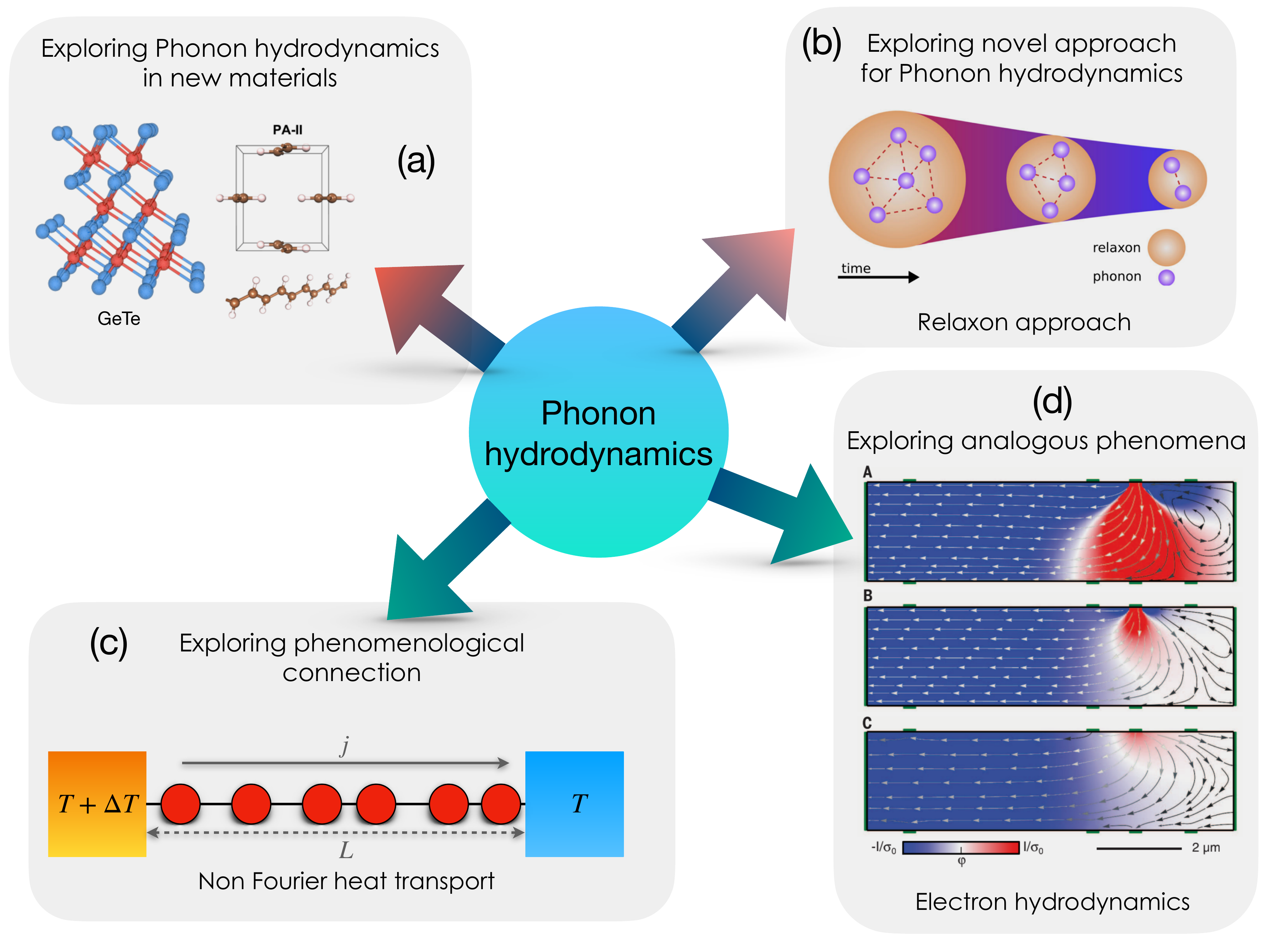}
    \caption{Different possible future research directions emerging from phonon hydrodynamics, as described in this review. (a) Exploring phonon hydrodynamics in new materials. Structure of GeTe: Reprinted (figure) with permission from Ref. \cite{kanka1}. Copyright (2020) American Physical Society. Structure of PA-II: Reprinted (figure) with permission from Ref. \cite{ZhangPolymer2020}. Copyright (2020) American Physical Society. (b) Exploring novel approaches to understand phonon hydrodynamics with more conceptual clarity. Relaxon schematic: Reprinted with permission from Ref. \cite{Cepellotti2016}. \href{https://creativecommons.org/licenses/by/3.0/}{CC BY 3.0}. (c) Exploring and establishing phenomenological connections with other non Fourier heat transport phenomena for better physical understanding of the subject. (d) Exploring similar phenomena in solids (e.g. electron hydrodynamics). Electron hydrodynamics: Reprinted (figure) with permission from Ref. \cite{BanduringrapheneEhydro}, AAAS. Republished with permission of [``CCC"], from \cite{BanduringrapheneEhydro}; permission conveyed through Copyright Clearance Center,Inc.}
    \label{fig:outlook}
\end{figure}

Graphene and other 2D materials show exceptional phonon hydrodynamics even at room temperature which is why most of the studies, dealing with novel applications on thermal transport, had been centered around these few materials. Nevertheless, the experimental, theoretical and numerical investigations opened up various other interesting physical and material related consequences in the realm of phonon physics and several new research pathways can be directed from the current understanding of the phonon hydrodynamics. We briefly discuss these possibilities here. Figure \ref{fig:outlook} summarizes the perspectives, outlook and different future research directions emerging out of the subject of phonon hydrodynamics.

\subsection{\textbf{Phonon hydrodynamics of organic systems}}

Recently, some of the organic materials are found \cite{ZhangPolymer2020} to display phonon hydrodynamics at moderate temperatures ($\approx$ 50 K for Polyacene and Polyacetylene and $\approx$ 120 K for Polyethylene). Crystalline polymers exhibit intrinsically bending acoustic modes due to the flexible property of the polymers. Possessing the flexural mode similar to that of the 2D graphene sheets enable these crystalline polymers to exhibit a strong anharmonicity and therefore to demonstrate phonon hydrodynamics. Weak van der Waals coupling between polymer chains are found to be responsible for the existence of the flexural phonon modes \cite{ZhangPolymer2020}. This investigation of phonon hydrodynamics in crystalline polymers will further help understanding the phonon transport in more complex organic systems. Moreover, consulting the controlling parameters, a new direction of research can be explored finding more complex materials with bending acoustic modes which seems to be a crucial property to enhance N scattering and overall three phonon scattering processes.

\subsection{\textbf{Unveiling the subtle phonon hydrodynamic features in materials}}

With the help of advanced experimental and first-principles techniques to solve Boltzmann transport equation, more subtle behavior of phonon hydrodynamics is seen to emerge from many crystalline materials that are conventionally unfit candidates for displaying phonon hydrodynamics. Recently, in the quest of experimentally observing the high frequency second sound in crystalline bulk Ge, Beardo $\textit{et al.}$ \cite{Beardo2021Ge} carried out a frequency-domain experiment using a rapidly varying temperature field and monitoring the phase lag of the thermal response of the material. Exploiting the second order time derivative of mesoscopic hyperbolic heat equation of the Maxwell, Cattaneo, and Vernotte type \cite{Cattaneo1948, Vernotte1958}, Beardo $\textit{et al.}$ \cite{Beardo2021Ge} employed an extremely high driving frequency to dominate the thermal inertial term over the damping term in the equation and therefore opening up a wide temperature window (7 K $<$ $T$ $<$ 300 K) to observe second sound feature in the high frequency limit in Ge which is otherwise isotopically enriched, resistive scattering dominated material. This is a significant experimental advancement in the field in terms of discovering phonon hydrodynamics in a more diverse pool of materials and therefore offers the scope of new physics and applications using those materials. In a series of separate recent $\textit{ab initio}$ density functional studies coupled with either solving complete Boltzmann transport equation or GK type hydrodynamic equation with Kinetic collective approach \cite{kanka2, kanka3}, subtle features of phonon hydrodynamics in crystalline GeTe, a low $\kappa$ material, had been evidenced. Both qualitative and quantitative numerical explorations using GK frequency conditions, second sound speed, thermal diffusivity, collective mode contributions, Knudsen number analysis were carried out and a fragile phonon hydrodynamic regime was identified \cite{kanka2} as a function of both temperature and characteristic size \cite{kanka3} of GeTe. Further the sensitivity of this regime with respect to the phonon scattering events corresponding to the isotopes and vacancies were understood. These studies lead not only to the possibilities to discover phonon hydrodynamics in other low $\kappa$ materials but also to the controlling and manipulating this feature. Therefore these experiments and numerical investigations are gradually opening up the prospects of accessing phonon hydrodynamics in more unexplored materials and helping in boosting the current understanding of the subject.

\subsection{\textbf{Going along the direction of the Relaxon approach}}

In a conceptually new method, the failure of the single mode relaxation time approach at low temperature and the advent of advanced computational strategies, helped in developing a complete solution of LBTE by taking into account the full scattering matrix \cite{Transport_waves_as_crystal_excitations, Cepellotti2016, Simoncelli2020}. This approach relies on representing collective excitations or `relaxon' as an eigenvalue equation with a measurable characteristic relaxation time. Relaxons can be thought of a linear combinations of single phonon excitations and thermal conductivity of materials can be realized using kinetic theory applied to the `relaxon' gas \cite{Cepellotti2016}. Apart from being computationally efficient, relaxon approach also brings forth new understanding of the collective excitations in materials. Phonon hydrodynamics was discovered and validated for graphene, graphite like materials \cite{Simoncelli2020, Cepellotti2016} via this method and a lot of new physics can be opened up in future using this novel approach if further employed on new materials.

\subsection{\textbf{Exploring electron hydrodynamics}}

Similar hydrodynamic behavior have recently been evidenced in the flow of electrons in strongly correlated systems \cite{BanduringrapheneEhydro, CrossnoEhydro, ZaanenEhydro, MollEhydro} where electronic transport is driven by the highly collective quantum states, giving a further hope to advance the electronic device technology. Electronic transport in graphene had been shown \cite{BanduringrapheneEhydro} to feature hydrodynamic behavior where electrons are seen to portray features similar to that of the viscous liquids. Though having a weak electron-phonon scattering, above the liquid Nitrogen temperature electron-electron scattering events are sufficiently frequent for graphene for local equilibrium and exhibiting viscous drags and hydrodynamics of electrons \cite{BanduringrapheneEhydro}. Moll $\textit{et al.}$ \cite{MollEhydro} experimentally found a significant viscous contribution to the resistance of the long conduction channels of metal PdCoO$_2$ of variable widths and therefore estimated electronic viscosity for PdCoO$_2$. Alike phonon hydrodynamics, electron hydrodynamics also occurs at specific conditions \cite{GangChenReview2021, ZaanenEhydro} and therefore this novel electronic feature also requires controlled nanofabrication for detection. Further, possibilities can emerge from a versatile material like graphene, which exhibits both strong phonon and electron hydrodynamic behavior, to harness exceptional thermoelectric properties by controlling and manipulating phonon-phonon, electron-electron and phonon-electron scattering.

\subsection{\textbf{Generic phenomenological connection with other studies on anomalous heat transport}}

Another pathway of research can proceed from understanding the phenomenological connections between phonon hydrodynamics and other non Fourier anomalous heat transport phenomena realized via various exactly solvable theoretical models. Especially, the analytical as well as large scale numerical simulations of the heat transport in Fermi-Pasta-Ulam (FPU) chains (also known as Fermi-Pasta-Ulam-Tsingou model or FPUT model) \cite{dhar2019fP, Suman2014PRE} and other 1D systems were found to show anomalous and non Fourier heat transport features \cite{dhar2008} which can be useful to compare with the phonon hydrodynamics results of the experimentally or numerically studied 1D systems like SWCNTs. The solution of the 1D models (e.g. 1D diatomic hard particle gas model \cite{Cipriani2005}) also indicated a superdiffusive spreading of the heat pulses \cite{Cipriani2005, dhar2019fP} if a heat pulse is introduced into the system. Nonlinear hydrodynamic fluctuation theory \cite{Spohn2016} suggested a Levy walk model to describe this superdiffusive heat transport, contrary to a random walk model which usually demonstrates the diffusive heat transport \cite{dhar2019fP}.

This Levy flight can be thought of as a drift motion of phonons where phonons collectively move in one direction for sufficiently long time steps before getting scattered. This is phenomenologically similar to the  central idea of phonon hydrodynamics, in which phonons perform a coherent drift motion at certain length and time scales. Of course, it is necessary to distinguish the proper size dependency of this drift to distinguish hydrodynamic with ballistic heat transport. Nevertheless, future research directions can be chalked out in creating much broader phenomenological links between various anomalous non Fourier heat transport phenomena.

\section{Acknowledgments}

This project has received funding from the European Union’s Horizon 2020 research and innovation program under Grant Agreement No. 824957 (“BeforeHand:” Boosting Performance of Phase Change Devices by Hetero- and Nanostructure Material Design).

\providecommand{\noopsort}[1]{}\providecommand{\singleletter}[1]{#1}%
%



\end{document}